\documentclass[11pt]{article}

\usepackage[textwidth=15.5cm,textheight=22.5cm]{geometry}
\usepackage{amsmath,amssymb}
\usepackage{latexsym}
\usepackage{graphicx}
\usepackage{cite}
\usepackage{bbm}
\usepackage{bm}
\usepackage{upgreek}
\usepackage{mathrsfs}

\newcommand{\be}{\begin{equation}}
\newcommand{\ee}{\end{equation}}
\newcommand{\ndt}{\noindent}
\def\bea{\begin{eqnarray}}
\def\eea{\end{eqnarray}}
\def\beas{\begin{eqnarray*}}
\def\eeas{\end{eqnarray*}}
\def\sla{\raise.15ex\hbox{$/$}\kern-.57em}

\def\parm{{\partial}_{-}}
\def\parp{\partial_+}

\newcommand\fr[1]{\frac{1}{#1}}

\newcommand {\nn} {\nonumber}
\def \ttb {\left(\theta\bar\theta\right)}

\renewcommand{\th}{\theta}
\newcommand{\bth}{\bar\theta}

\allowdisplaybreaks[1]

\newcommand{\er}{{\rm e}}
\newcommand{\dr}{{\rm d}}
\newcommand{\Tr}{{\rm Tr}}

\newcommand{\Z}{\mathbbm{Z}}

\newcommand{\del}{\partial}
\newcommand{\ba}{\begin{eqnarray}}
\newcommand{\ea}{\end{eqnarray}}
\newcommand{\bdm}{\begin{displaymath}}
\newcommand{\edm}{\end{displaymath}}
\newcommand{\ra}{\rangle}
\newcommand{\la}{\langle}

\newcommand{\one}{\mathbbm{1}}

\def\S{\Sigma}

\def\b{\beta}
\def\a{\alpha}
\def\g{\gamma}
\def\s{\sigma}

\def\veps{\varepsilon}

\def\gdot{{\dot\gamma}}
\def\adot{{\dot\alpha}}
\def\bdot{{\dot\beta}}
\def\d{\delta}

\def\v{\varphi}

\newcommand{\half}{\frac{1}{2}}

\newcommand{\ie}{{\it i.e.\ }}
\newcommand{\eg}{{\it e.g.\ }}
\newcommand{\wtilde}{\widetilde}
\newcommand{\bmhv}{$\overline{\rm MHV}$}
\newcommand{\calA}{{\mathcal A}}

\newcommand{\calL}{{\mathcal L}}

\newcommand{\calN}{{\mathcal N}}
\newcommand{\calO}{{\mathcal O}}
\newcommand{\calP}{{\mathcal P}}

\newcommand{\calS}{{\mathcal S}}

\DeclareMathAlphabet{\mathpzc}{OT1}{pzc}{m}{it}

\newcommand\hsp[1]{\hspace*{#1 cm}}
\newcommand\vsp[1]{\vspace*{#1 cm}}

\newcommand{\bfl}{\bm{l}}

\newcommand{\bfp}{\bm{p}}
\newcommand{\bfq}{\bm{q}}
\newcommand{\bfr}{\bm{r}}

\begin{document}

\begin{titlepage}
\begin{flushright}    
{\small DIAS-STP-11-01}
\end{flushright}

\vsp{1.5}
\centerline{\LARGE {\bf {A manifestly MHV Lagrangian for $\mathcal N=4$ 
Yang--Mills}}}

\vskip 1cm

\centerline{Sudarshan Ananth$^\dagger$, Stefano Kovacs$^*$ 
and Sarthak Parikh$^\dagger$}

\vskip .5cm

\centerline{{\it {$^\dagger$ Indian Institute of Science Education and
Research}}} \centerline{{\it {$\;$ Pune 411021, India}}}

\vskip 0.5cm

\centerline{\em $^*$ Dublin Institute for Advanced Studies}
\centerline{\em $\;$ 10 Burlington Road, Dublin 4, Ireland}

\vskip 1cm

\centerline{\bf {Abstract}}

\vskip .5cm

\noindent 
We derive a manifestly MHV Lagrangian for the $\mathcal N=4$
supersymmetric Yang--Mills theory in light-cone superspace. This is
achieved by constructing a canonical redefinition which maps the
$\calN=4$ superfield, $\phi$, and its conjugate, $\bar\phi$, to a new
pair of superfields, $\chi$ and $\wtilde\chi$. In terms of these new
superfields the $\calN=4$ Lagrangian takes a (non-polynomial)
manifestly MHV form, containing vertices involving two superfields of
negative helicity and an arbitrary number of superfields of positive
helicity. We also discuss constraints satisfied by the new
superfields, which ensure that they describe the correct degrees of
freedom in the $\calN=4$ supermultiplet. We test our derivation by
showing that an expansion of our superspace Lagrangian in component
fields reproduces the correct gluon MHV vertices. 

\end{titlepage}

\section{Introduction}

The $\calN=4$ supersymmetric Yang--Mills (SYM) theory has a number of
remarkable properties, some of which have been known for a long time
and  others which have emerged more recently. The theory has maximal
rigid supersymmetry and is an example of an interacting conformal
field theory in four dimensions. It has been extensively studied --
together with some of its deformations -- for the special role it
plays in the AdS/CFT correspondence~\cite{Madscft,GKP,Wadscft}. The
original formulation of the correspondence relates $\calN=4$ SYM to
type IIB string theory in an AdS$_5\times S^5$ background and this
remains the best understood and most thoroughly tested example of the
duality. The $\calN=4$ Yang--Mills theory also possesses a
non-perturbative SL(2,$\Z$) symmetry, known as S-duality, which
generalises the electric-magnetic duality of Maxwell's equations in
the vacuum. This symmetry has recently been studied in connection with
the so-called geometric Langlands program~\cite{KW,GW,EW2}. In another
remarkable development, an integrable structure underlying the
spectrum of scaling dimensions of $\calN=4$ gauge-invariant composite
operators has been discovered, see~\cite{integra} and references
therein for a recent comprehensive review. This finding is a further
indication of the richness of the $\calN=4$ theory. It is also the
basis of powerful techniques developed for the calculation of quantum
corrections to the scaling dimensions, which have allowed extremely
accurate tests of the AdS/CFT duality~\cite{BHJL}. 

In the past few years the study of scattering amplitudes in the
$\calN=4$ theory has attracted considerable interest. This is in spite
of the fact that, as a conformally invariant non-Abelian gauge theory,
$\calN=4$ SYM does not possess well defined asymptotic states, making
the physical relevance of scattering amplitudes somewhat dubious. It
is, however, useful to separate the issue of the physical
interpretation of the scattering amplitudes from their mathematical
properties. Scattering amplitudes in the $\calN=4$ Yang--Mills theory
are formally well defined: they are affected by infra-red divergences
-- as is normally the case in theories with massless particles --
which, however, can be dealt with in a standard way and, moreover,
they are free of ultra-violet divergences. If one does not insist on
attributing to them a direct physical meaning, the $\calN=4$ scattering
amplitudes have a number of interesting properties. They share some of
the features of amplitudes in phenomenologically more interesting
theories and moreover possess some unique and intriguing
properties. The need to better understand such properties and their
implications provides the main motivation for this paper. Our focus is
on constructing a formalism which may be useful in this respect,
rather than on developing more efficient computational tools. 

It is useful to consider $\calN=4$ SYM in the more general context of
the study of scattering amplitudes in non-Abelian gauge theories.
Irrespective of details such as the exact matter content or the amount
of supersymmetry, perturbative scattering amplitudes in Yang--Mills
theories possess remarkable features, which are not easily explained
within the framework of a standard Lagrangian formulation. Among such
features, the most striking is the simplicity of tree-level and, to a
lesser extent,  one-loop amplitudes.  This is particularly evident
when one considers planar amplitudes with external states of definite
helicity and focusses on the so-called colour-ordered partial
amplitudes, as opposed to full cross-sections. A $n$-gluon
amplitude~\footnote{For simplicity, in this general discussion we
focus on gluon amplitudes.}, $\calA_n$, can be decomposed into a sum
of the form 
\be
\calA_n(\{p_i,h_i, a_i\}) = (2\pi)^4 \delta^{(4)}\!\!
\left(\sum_{i=1}^n p_i\right) \!\!\sum_{\s\in S_n/\Z_n} \!\!g^{n-2}
\Tr\left(T^{a_{\s(1)}} \cdots T^{a_{\s(n)}}\right)
A_n\left(\s(p_1,h_1;\ldots;p_n,h_n)\right) , 
\ee 
where $p_i$, $i=1,\ldots,n$, are the null on-shell momenta of the
external particles, the $h_i$'s are their helicities and the $a_i$'s
denote their colour indices. The sum is over all non-cyclic
permutations of the labels. The $A_n$'s on the right hand side are the
colour-ordered partial amplitudes. They depend only on the momenta and
helicities of the $n$ gluons and we use the simplified  notation 
\be
A_n(p_1,h_1;\ldots;p_n,h_n) \equiv A_n(1^{h_1},\ldots,n^{h_n}) \, ,
\label{part-ampl}
\ee
with $h_i=\pm$. 

A first notable result pertaining to helicity partial amplitudes such
as (\ref{part-ampl}) is the existence of unexpected selection
rules~\cite{all+}. Amplitudes with all $n$ gluons of the same
helicity~\footnote{We adopt the convention that all the particles in a
scattering amplitude are incoming.} and those with $n-1$ gluons of one
helicity and a single gluon of opposite helicity vanish at tree
level. The same amplitudes are zero to all orders in supersymmetric
theories. The simplest class of non-trivial amplitudes consists of
those with two gluons of one helicity and all the others of the
opposite helicity. Amplitudes of this type with two gluons of negative
helicity and $n-2$ gluons of positive helicity are referred to as
maximally helicity violating (MHV). In the conventional terminology
next-to-MHV amplitudes, denoted by NMHV, are those with three negative
helicities. In general N$^k$MHV are amplitudes of the type
$A_n(1^-,2^-,\ldots,(k+2)^-,(k+3)^+,\ldots,n^+)$, plus all possible
permutations of the momenta, in which $k+2$ gluons have negative
helicity. Anti-MHV amplitudes, denoted by \bmhv, are those with two
positive and an arbitrary number of negative helicity gluons. 

The colour-ordered $n$-gluon partial amplitude in the MHV case (for
arbitrary $n$) has an extremely simple and elegant form, which was
first conjectured in~\cite{PT} and later derived in~\cite{BG}. A
generating function for $\calN=4$ MHV super-amplitudes was obtained
in\cite{VPN}. The remarkable simplicity of the MHV partial amplitudes
in Yang--Mills theory is completely obscured in a standard Lagrangian
formulation. The calculation of even the simplest (MHV) tree
amplitudes using traditional methods based on Feynman rules derived
from a Lagrangian is very tedious and becomes formidably complicated
as the number of external particles grows~\footnote{To give an idea of
the complexity of tree level calculations based on Feynman rules we
recall that for a ten-gluon amplitude the number of diagrams
contributing is of the order of $10^7$~\cite{MP}}. 

Motivated by the need to overcome the cumbersome nature of the
techniques  based on Feynman diagrams,  much work has been done over
the past two decades to develop and refine more efficient methods for
the calculation of perturbative scattering amplitudes in non-Abelian
gauge theories, see~\cite{BDetal} for reviews. A particularly
interesting proposal was presented in~\cite{CSW} in the form of
so-called ``MHV rules''. These authors, inspired by ideas from
topological string theory in twistor space~\cite{EW3}, proposed a
novel formalism for the evaluation of tree level scattering
amplitudes, which uses as building blocks vertices consisting of a
certain off-shell continuation of the simple MHV amplitudes.
According to this proposal, amplitudes with generic external
helicities are obtained by sewing together (off-shell) MHV amplitudes
using scalar propagators. The approach based on these MHV rules
significantly reduces the complexity of the calculation of non-MHV
tree-level amplitudes.  A proof of the MHV rules was given
in~\cite{KR} using recursion relations for tree-level
amplitudes~\cite{BCF+W}. The approach based on the MHV rules has been
successfully extended to loop level. In particular, one-loop MHV
amplitudes have been computed in~\cite{BST}, reproducing the results
previously obtained using unitarity methods~\cite{BDDK}. 

In a subsequent development a ``MHV Lagrangian'', which generates the
MHV rules for pure Yang--Mills theory, was constructed
in~\cite{GR,PM}. As presented in~\cite{PM}, the new Lagrangian is
obtained from the ordinary Yang--Mills Lagrangian through a canonical
redefinition of the fields. The result is a non-polynomial Lagrangian
in which the vertex of order $n$ is directly related to the $n$-gluon
MHV amplitude. Of course, being obtained via a field redefinition from
the original Lagrangian, the one obtained in~\cite{PM} provides a new,
but in every respect equivalent, description of Yang--Mills
theory. This means that the MHV Lagrangian is suitable for the
calculation of scattering amplitude, both at tree and at loop level,
but also of other quantities such as correlation functions. In this
paper we present the derivation of a manifestly MHV Lagrangian for
$\calN=4$ SYM in light-cone superspace. Our construction, both in
intermediate steps and in the final result, follows
closely~\cite{PM,EM}. We will obtain a Lagrangian containing an
infinite series of vertices, each involving two superfields of
helicity $-1$ and a number of superfields of helicity $+1$. 

A MHV Lagrangian in light-cone superspace for $\calN=4$ SYM was
previously constructed in~\cite{FH}. There are some fundamental
differences in the approach that we advocate here compared to that of
that paper. The Lagrangian that we obtain is also different, although
our result and that of~\cite{FH} agree when expanded in component
fields.  The central element which distinguishes our approach from
that of~\cite{FH} is that we work with a pair of constrained
superfields, of helicity $+1$ and $-1$ respectively, while the authors
of~\cite{FH} solve explicitly the constraints to write the Lagrangian
in terms of a single superfield.  Our approach leads to additional
subtleties -- which we will analyse in detail -- but allows us to
construct a Lagrangian which is {\it manifestly} MHV in superspace. We
will further discuss similarities and differences between our
derivation and that in~\cite{FH} in the concluding section. 

While the main application of the MHV Lagrangian in the pure
Yang--Mills case is the calculation of actual physical amplitudes, we
believe that the $\calN=4$ construction presented here will be useful
in order to better understand some of the structures which arise in
the study of scattering amplitudes, but may have more general
relevance beyond that specific application.

From a practical point of view the study of scattering amplitudes in
the $\calN=4$ theory can be viewed as a testing ground to develop
computational techniques in a setting where the complications
associated with ultra-violet divergences are not present. However,
viewed as formal objects, the $\calN=4$ amplitudes display some
peculiar features which are unique to this theory and interesting in
their own right. 

Among the remarkable properties of the $\calN=4$ scattering amplitudes
is a duality relating them to certain polygonal Wilson loops. This
duality asserts that the MHV colour ordered partial amplitudes
(\ref{part-ampl}) are related to the expectation value of certain
Wilson loops defined on polygonal contours. A more precise formulation
of the duality can be phrased as follows. The exact planar MHV
$n$-point partial amplitude can be factorised as
\be
A^{\rm (MHV)}_n(p_1,\ldots,p_n) = A^{\rm (MHV)}_{n,{\rm tree}} 
{\hat A}_n(p_1,\ldots,p_n) \, ,
\label{factor-ampl}
\ee
where $A^{\rm (MHV)}_{n,{\rm tree}}$ is the tree-level MHV
amplitude~\cite{PT,BG}. According to the duality, the factor $\hat
A_n(p_,\ldots,p_n)$ in (\ref{factor-ampl}) is to be identified with
the expectation value of a Wilson loop defined by the polygon of
vertices $x_i$, $i=1,\ldots,n+1$, with  
\be 
x_{i+1}-x_i = p_i, \quad
i=1,\ldots,n \quad {\rm and} \quad x_{n+1}=x_1 \, .
\label{dualvariables}
\ee
Therefore the statement is
\be
\hat A_n(p_1,\ldots,p_n) = \la W(x_1,\ldots,x_n) \ra \, .
\ee
This correspondence was originally proposed in~\cite{AM} as a means of
computing the strong coupling limit of scattering amplitudes via the
AdS/CFT correspondence. However, it has subsequently been tested as a
perturbative duality of the $\calN=4$ Yang--Mills theory without any
reference to the dual gravitational description~\cite{W-sa-pert}.  

This duality is intimately related to one of the most intriguing
features of the  $\calN=4$ theory, a recently discovered novel
symmetry, referred to as dual superconformal symmetry, displayed by
planar scattering amplitudes~\cite{DHKS,BM}. The $\calN=4$ SYM theory
is classically invariant under the superconformal group PSU(2,2$|$4)
and this symmetry remains unbroken to all orders in perturbation
theory~\cite{N4fin,SM,BLN2}. The PSU(2,2$|$4) supergroup contains
SO(2,4)$\times$SO(6)$_R$$\,\sim\,$SU(2,2)$\times$SU(4)$_R$ as maximal
bosonic subgroup, where SO(2,4)$\,\sim\,$SU(2,2) is the
four-dimensional conformal group and SO(6)$_R\,\sim\,$SU(4)$_R$ is the
$\calN=4$ R-symmetry. It has been observed that scattering amplitudes
in $\calN=4$ SYM, when expressed in terms of suitable auxiliary
variables~\footnote{The auxiliary variables are precisely the
positions, $x_i$, introduced in (\ref{dualvariables}).}, possess an
additional PSU(2,2$|$4) symmetry, which is not related to the original
superconformal symmetry in any obvious way. It has recently been
noted that ordinary and dual superconformal symmetry algebras
appear to combine to generate a Yangian symmetry of the type
encountered in the study integrable systems~\cite{DHP}. It has been
speculated that this infinite dimensional symmetry may allow to
completely determine the $\calN=4$ S-matrix through purely algebraic
means.

A clear understanding of the origin of this new symmetry and of its
relevance beyond the study of scattering amplitudes is still
lacking. The fact  that the dual superconformal symmetry appears to be
present only in the planar approximation makes it difficult to
directly trace it to characteristics of the Lagrangian. It seems,
however, unlikely that this symmetry can be an accidental property
uniquely seen in scattering amplitudes. Recent work relating
amplitudes and null polygonal Wilson loops to special limits of
correlation functions~\cite{corr-Wloop,corr-ampl} offers interesting
insights into these issues. One of the aims of the present paper is to
develop a formalism which may help to shed light on the structure and
implications of the dual superconformal symmetry. The reformulation of
the $\calN=4$ theory that we present appears to be well suited for
this purpose, since it  has built-in some of the basic features of
scattering amplitudes and moreover it is manifestly $\calN=4$
supersymmetric. We intend to pursue this line of investigation in the
future. More generally, it will be interesting to use this new MHV
formulation to revisit other aspects of the $\calN=4$ SYM theory such
as ultra-violet finiteness and possible connections with $\calN=8$
supergravity.

This paper is organised as follows. In section \ref{N4-LCSS} we review
the formulation of $\calN=4$ SYM in light-cone superspace, with
particular emphasis on the aspects which are relevant for the
discussion of scattering amplitudes. In section \ref{SFredef} we
construct a canonical change of variables which yields the new
superfields used in the MHV Lagrangian. The explicit form of the
leading terms in this Lagrangian is presented in section
\ref{sec:MHVlag}. In section \ref{components} we discuss the form of
our Lagrangian in terms of component fields and we show that it
reproduces the known terms in the MHV Lagrangian for the pure
Yang--Mills case. Various technical details are discussed in the
appendices.

\section{$\mathcal N=4$ Yang--Mills in light-cone superspace}
\label{N4-LCSS}

In this section we briefly review the formulation of $\calN=4$ SYM in
light-cone superspace. This formalism provides a description of the
theory, in terms of the sole physical degrees of freedom, in which the
full $\calN=4$ supersymmetry as well as the SU(4)$_R$ R-symmetry are
manifest. We also discuss the helicity assignments for the various
fields in the theory both in components and in superspace. In the
following sections this formulation of the theory will allow us to
identify a field redefinition which brings the action into a
manifestly MHV form, along the lines of the construction proposed in
\cite{GR,PM,EM} for the pure Yang--Mills case. 

\subsection{Light-front quantisation} 

We work with space-time signature $(-,+,+,+)$ and we choose a null
unit vector $\zeta_\mu$, which identifies the time direction used in
the light-front quantisation. With the choice $\zeta_\mu =
\fr{\sqrt{2}}(+1,0,0,+1)$, the light-cone coordinates and their
derivatives are
\bea
&& \hsp{-1}x^{\pm} = \frac{1}{\sqrt 2}({x^0}\,{\pm}\,{x^3})\, , \quad
x = \frac{1}{\sqrt 2}({x^1}+i\,{x^2}) \, , \quad 
{\bar x} = \frac{1}{\sqrt 2}({x^1}-i\,{x^2})\, , 
\label{lc-ccord} \\
&& \hsp{-1} {\partial_{\pm}}=\frac{1}{\sqrt 2}({\partial_0}\,{\pm}\,
{\partial_3})\,, \quad {\bar\partial} =\frac{1}{\sqrt 2}
({\partial_1}-i\,{\partial_2})\, , \quad
{\partial} =\frac{1}{\sqrt 2}({\partial_1}+i\,{\partial_2})\, .
\label{lc-deriv}
\eea
and $x^+=\zeta_\mu x^\mu$ will be taken as light-cone time. 

In keeping with the literature, we represent momentum vectors as
bi-spinors, mapping $p_\mu$ to $p_{\a\adot} =
\sigma^\mu_{\a\adot}p_\mu$, where $\s^\mu = (-\one,\s^i)$ and $\s^i$, 
$i=1,2,3$, are Pauli matrices. In terms of the light-cone components 
of $p_\mu$ we have
\be
p_{\a\adot} = \sqrt{2}\left(\!\!\begin{array}{cc} -p_- & \bar p \\ 
p & -p_+ \end{array}\!\!\right) \, .
\ee
A light-like vector, $p_\mu$, can be written as 
\be
p_{\a\adot} = \lambda_\a\wtilde\lambda_\adot \, ,
\label{spinormomentum}
\ee
for (commuting) spinors $\lambda_\a$ and $\wtilde\lambda_\adot$ of
positive and negative chirality respectively. For $p_{\a\adot}$ to be real 
one must take $\wtilde\lambda=\pm\bar\lambda$. We can choose for 
instance
\be
\lambda_\a = 2^{1/4} \left(\!\!\begin{array}{c} \sqrt{p_-} \\
\displaystyle -\frac{p}{\sqrt{p_-}} \rule{0pt}{19pt}\end{array} \!\! 
\right) \quad {\rm and} 
\quad \wtilde\lambda_\adot = 2^{1/4} \left( \!\!\begin{array}{c} 
-\sqrt{p_-} \\ \displaystyle\frac{\bar p}{\sqrt{p_-}} \rule{0pt}{19pt} 
\end{array}\!\! \right) \, .
\ee
From the light-like vector $\zeta_\mu$ we construct $\zeta_{\a\adot} =
\s^\mu_{\a\adot}\zeta_\mu$, which we represent as $\zeta_{\a\adot} =
\nu_\a\wtilde\nu_\adot$, with 
\be
\nu_\a = \left(\!\!\begin{array}{c} 0 \\ 2^{1/4} \end{array} \!\!\right) 
\quad {\rm and} \quad \wtilde\nu_\adot = 
\left(\!\!\begin{array}{c} 0 \\ -2^{1/4} 
\end{array} \!\!\right) \, .
\label{zetaspinors}
\ee  
Given two spinors of positive chirality, $\lambda_\a$ and $\mu_\a$, we
can construct a Lorentz invariant bilinear, 
\be
\la \lambda\,\mu\ra = \veps_{\a\b}\lambda^\a\mu^\b \, ,
\label{abracket}
\ee
where the tensor $\veps_{\a\b}$ used to lower indices is
anti-symmetric with $\veps_{12}=1$ and its inverse is $\veps^{\a\b}$
satisfying $\veps_{\a\b}\veps^{\b\g}=\d_\a^\g$. Similarly out of two
negative chirality spinors, $\wtilde\lambda_\adot$ and
$\wtilde\mu_\adot$, we define the invariant bilinear 
\be
[\wtilde\lambda \, \wtilde\mu] = \veps_{\adot\bdot} \lambda^\adot 
\mu^\bdot \, ,
\label{sqbracket}
\ee
where the anti-symmetric tensor $\veps_{\adot\bdot}$ is defined
similarly to $\veps_{\a\b}$. Its inverse is $\veps^{\adot\bdot}$ and
they satisfy $\veps_{\adot\bdot}\veps^{\bdot\gdot}=\d_\adot^\gdot$. We
will make use of the two invariant products (\ref{abracket}) and
(\ref{sqbracket}) in the discussion of helicity.

\subsection{$\calN=4$ SYM in light-cone superspace}
\label{N4LCsuperspace}

The field content of the $\calN=4$ SYM theory comprises a gauge field,
$A_\mu$, four Weyl fermions, $\psi^m_\a$, and their conjugates,
$\bar\psi_{m \,\adot}$, $m=1,\ldots,4$, and six real scalars, $\v^i$,
$i=1,\ldots,6$. The gauge field is a SU(4)$_R$ singlet, the fermions
transform in the $\mathbf{4}$ and $\mathbf{\bar 4}$ and the scalars in
the $\mathbf{6}$. The gauge field components are
\be
A_\pm = \fr{\sqrt{2}}(A_0\pm A_3) \, , \quad A = \fr{\sqrt{2}} 
( A_1+ i A_2)\, , \quad \bar A = \fr{\sqrt{2}} ( A_1 - i A_2) \, .
\ee
From $A_\mu$ we can construct $A_{\a\adot} = \sigma^\mu_{\a\adot}
A_\mu$. In terms of light-cone components we get 
\be
A_{\a\adot} = \sqrt{2}\left(\!\!\begin{array}{cc} -A_- & \bar A \\ 
A & -A_+ \end{array}\!\!\right) \, .
\label{Aspinor}
\ee
The light-cone gauge description of the theory uses only physical
degrees of freedom. We fix the gauge setting $A_-=0$ and we integrate
out $A_+$, leaving the two transverse components, $A$ and $\bar
A$. Similarly the four Weyl fermions, $\psi^m_\a$, and their
conjugates, $\bar\psi_{m\,\adot}$, are decomposed according to the
projection
\be
\psi^m_\a ~ \to ~ \psi_{(\pm)}^m = \calP_\pm \psi^m_\a \, , \qquad
\bar\psi_{m\,\adot} ~ \to ~ \bar\psi^{(\pm)}_m = 
\calP_\pm \bar\psi_{m\,\adot} \, ,
\ee
where $\calP_\pm = -\fr{\sqrt{2}} \sigma^\pm$, with $\s^\pm = \fr{\sqrt{2}}
(\s^0\pm\s^3)$. We then integrate out the $\psi^m_{(+)}$ and
$\bar\psi_m^{(+)}$ components, leaving four one-component fermionic
fields and their conjugates,
\be
\lambda^m \equiv \psi^m_{(-)} \, , \qquad 
\bar\lambda_m \equiv \bar\psi^{(-)}_m \, .
\ee
The $\calN=4$ multiplet is completed by the six real scalar fields,
which we represent as SU(4)$_R$ bi-spinors, $\v^{mn}$,
$m,n=1,\ldots,4$, satisfying the reality condition
\be
\bar\v_{mn} \equiv \left(\v^{mn}\right)^* = \half \veps_{mnpq}\v^{pq} \, .
\ee
In the following it will be important that the physical fields used in
the light-cone gauge,
\be
(A,\bar A, \v^{mn}, \lambda^m, {\bar\lambda}_m) \, ,
\label{physfields}
\ee
can be assigned definite helicities.  
 
The $\calN=4$ light-cone superspace is made up of the four bosonic
coordinates (\ref{lc-ccord}) and eight fermionic coordinates,
$\theta^m$ and  $\bar\theta_m$, $m=1,\ldots,4$, transforming in the
$\mathbf{4}$ and $\mathbf{\bar 4}$ of SU(4)$_R$. When working in
configuration space we will collectively denote the superspace
coordinates by $z=(x^+,x^-,x,\bar x,\theta^m,\bar\theta_m)$. The full
$\calN=4$ supersymmetry is manifest, with half of the supercharges
(denoted by $q^m$ and $\bar q_m$ and referred to as kinematical)
realised linearly as translations in the fermionic coordinates and the
other half (referred to as dynamical) non-linearly realised.

We also introduce the superspace chiral derivatives, $d^m$ and ${\bar
d}_m$, defined as 
\be
d^m = -\frac{\partial}{\partial\bar\theta_m} 
+\frac{i}{\sqrt{2}} \theta^m\parm \,, \qquad 
\bar d_m = \frac{\partial}{\partial\theta^m} 
-\frac{i}{\sqrt{2}} \bar\theta_m\parm \, , \quad m=1,\ldots,4 \, .
\label{chiralder}
\ee
They obey
\be
\{d^m,{\bar d}_n\} = i\sqrt{2}\, \d^m_n \,\del_- \, 
\label{dsusyalg}
\ee
and anticommute with the supercharges $q^m$ and $\bar q_m$.

An irreducible representation of the $\calN=4$ super-algebra is
realised in terms of a single complex superfield,
$\phi(x,\theta,\bar\theta)$, which contains all the fields
(\ref{physfields}) as components. The superfield $\phi(x,\th,\bth)$ is
a SU(4)$_R$ singlet defined by the constraints~\cite{BLN1,BLN2}
\be
d^m \phi(x,\theta,\bar\theta) = 0 \, , \qquad 
{\bar d}_m {\bar d}_n \phi(x,\theta,\bar\theta) = \half \veps_{mnpq}
d^p d^q \bar\phi(x,\theta,\bar\theta) \, ,
\label{origconstraints}
\ee
where $\bar\phi = \phi^*$ satisfies ${\bar d}_m
\bar\phi(x,\theta,\bar\theta) =0$. The unique solution to these
constraints is a superfield with the following component 
expansion~\cite{BLN1}
\bea
\phi\,(x,\theta,\bar\theta)&\!\!=\!\!&-\frac{1}{\parm}A(y)
-\frac{i}{\parm}\theta^m{\bar \lambda}_m(y)
+\frac{i}{\sqrt 2}\,\theta^m\theta^n{\bar \v}_{mn}(y)\nn \\
&&+\frac{\sqrt 2}{6}\theta^m\theta^n\theta^p\veps_{mnpq}\lambda^q(y) 
-\frac{1}{12}\,\theta^m\theta^n\theta^p\theta^q
\veps_{mnpq}\parm{\bar A}(y) \, ,
\label{N4superfield}
\eea
where we introduced the chiral variable 
\be
y=(x^+,y^-=x^--\frac{i}{\sqrt{2}}\theta^m\bar\theta_m,x,\bar x) 
\label{chirvar}
\ee
and the right hand side is understood to be a power expansion about
$x^-$. Appendix \ref{more-on-constraints} contains a more detailed
discussion of constraints in light-cone superspace.  

In terms of the superfields $\phi$ and $\bar\phi$, the $\mathcal N=4$ SYM
light-cone action is~\cite{SM,BLN1}
\be
\label{n=4}
\calS = 72 \int \dr^4x \int \dr^4\theta\,\dr^4 \bar \theta \: 
{\cal L}(\phi,\bar\phi,\del_\mu\phi,\del_\mu\bar\phi)\ ,
\ee
where the Lagrangian density,
$\calL(\phi,\bar\phi,\del_\mu\phi,\del_\mu\bar\phi)\equiv
\calL_{\phi,\bar\phi}$, is
\bea
{\cal L}_{\phi,\bar\phi} &\!\!=\!\!&\Tr\left\{ -2\,\bar \phi\, 
\frac{\square}{\parm^2}\, \phi + i\frac{8}{3}g \left( \fr{\parm}
\bar\phi\, \left[\phi,\bar\partial \phi \right] + \fr{\parm} \phi\, 
\left[ \bar\phi,\partial \bar\phi\,\right] \right) \right.\nn \\
 && \left. \hsp{0.7}+ 2g^2 \left(\fr{\parm} \left[\phi,\parm \phi\right] 
\fr{\parm} \left[\bar \phi,\parm \bar \phi\,\right] + \frac{1}{2}
\left[\phi,\bar \phi\,\right]\left[\phi,\bar \phi\,\right] \right) \right\} ,
\label{Lagradensity}
\eea
and the d'Alembertian in light-cone coordinates reads
\be
\Box=2(\partial{\bar \partial}-\parp\parm)\, .
\ee
The Grassmann integrations are normalised so that 
\bdm
\int \dr^4\theta\,\theta^1\theta^2\theta^3\theta^4=1 \, .
\edm
Notice also that here and in the following we use the prescription of
\cite{SM} for the $\frac{1}{\del_-}$ operator.

\subsection{Helicity} 
\label{helicity}

Scattering amplitudes in a massless theory are traditionally computed
as functions of the momenta -- which enter through invariant
combinations such as the Mandelstam variables -- and polarisation
vectors of the incoming and outgoing particles. However, it has been
observed \cite{BDetal} that sub-amplitudes in which the external
states have definite helicities are considerably simpler than the full
amplitudes with arbitrary external momenta and polarisations. The
relative simplicity of these helicity amplitudes is at the heart of
the development of efficient techniques for the calculation of
scattering amplitudes, which has seen remarkable progress in the past
decade. The light-cone description provides the natural framework to
analyse scattering amplitudes in which the external states have
definite helicity.

In order to describe states of given light-like momentum and helicity
it is convenient to work with the representation
(\ref{spinormomentum}). The two spinors $\lambda_\a$ and
$\wtilde\lambda_\adot$ -- together with
$\zeta_{\a\adot}=\nu_\a\wtilde\nu_\adot$ defining the time direction
in light-front quantisation -- determine both the momentum and the
helicity of a massless state. Starting with the spinors $\lambda_\a$
and $\wtilde\lambda_\adot$ describing the light-like momentum
$p_{\a\adot}=\lambda_\a\wtilde\lambda_\adot$, we can construct
polarisation vectors corresponding to positive and negative helicity
states. A positive helicity polarisation vector is obtained as
\be
\eta^{(+)}_{\a\adot} = \frac{\nu_\a\wtilde\lambda_\adot}
{[\wtilde\nu\,\wtilde\lambda]} \, ,
\label{+helicity}
\ee
with $\nu_\a$ and $\wtilde\nu_\adot$ defined in (\ref{zetaspinors}). A
negative helicity polarisation vector is
\be
\eta^{(-)}_{\a\adot} = \frac{\lambda_\a\wtilde\nu_\adot}
{\la\nu\,\lambda\ra} \, .
\label{-helicity}
\ee
The vectors $\eta^{(+)}_\mu$ and $\eta^{(-)}_\mu$ thus defined satisfy
$\eta^{(+)}_\mu p^\mu = \eta^{(-)}_\mu p^\mu = 0$, as required for
polarisation vectors, thanks to the identities 
\be
\la\lambda \, \lambda \ra = 0 \, , \qquad [\wtilde\lambda \, 
\wtilde\lambda] = 0 \, .
\ee
Comparing the explicit form of (\ref{+helicity}) and (\ref{-helicity}),
\be
\eta^{(+)}_{\a\adot} = \left(\!\begin{array}{cc} 0 & 0 \\ 
1 & \displaystyle -\frac{\bar p}{p_-} \rule{0pt}{15pt}
\end{array}\! \right) \, , \qquad \eta^{(-)}_{\a\adot} = 
\left( \! \begin{array}{cc} 0 & 1 \\ 0 & \displaystyle 
-\frac{p}{p_-} \rule{0pt}{15pt} \end{array} \! \right) \, ,
\ee
with the spinor representation (\ref{Aspinor}) of $A_\mu$ in
light-cone gauge, we conclude that the physical components of the
gauge field, $A$ and $\bar A$, describe gluons of positive and
negative helicity respectively.  Similarly one can show that the two
fermionic fields, $\lambda^m$ and $\bar\lambda_m$, describe gauginos
of $-1/2$ and $+1/2$ helicity respectively. 

The above analysis reflects the fact that in the light-cone gauge we
can identify helicity with the U(1) charge associated with rotations
in the transverse $(x,\bar x)$ plane. Complex fields are used to
describe particles with helicity. Real fields describe helicity zero
particles (Lorentz scalars). In the case of $\calN=4$ SYM the complex
field $A$ has U(1) charge $+1$ and describes gluons of positive
helicity, its complex conjugate, $\bar A$, has U(1) charge $-1$ and
describes negative helicity gluons. Similarly $\lambda^m$ and its
conjugate $\bar\lambda_m$ have U(1) charge $-1/2$ and $+1/2$; they
describe gauginos of negative and positive  helicity respectively. The
six scalars in the theory are described by real fields, $\v^{mn}$,
which are not charged under U(1)~\footnote{The U(1) charges of the
fields in the $\calN=4$ SYM multiplet can be understood recalling that
the theory is the dimensional reduction of $\calN=1$ SYM in ten
dimensions. In the ten-dimensional light-cone formulation the
$\calN=1$ theory has SO(8) invariance. The field content consists of a
vector and a Majorana--Weyl spinor, transforming in the
representations $\bf{8}_{\rm v}$ and $\bf{8}_{\rm s}$ of SO(8). Upon
dimensional reduction to $d=4$ we consider the decomposition
SO(8)$\,\supset\,$SO(2)$\times$SO(6)$_R\sim\,$U(1)$\times$SU(4)$_R$,
where SO(2)$\,\sim\,$U(1) corresponds to rotations in the transverse
$(x,\bar x)$ directions and SO(6)$_R\sim\,$SU(4)$_R$ is the $\calN=4$
R-symmetry. The branching rules for the decomposition give (the
subscripts denote the U(1) charge)
\[
\bf{8}_{\rm v} = \bf{6}_0 \oplus \bf{1}_{+1} \oplus \bf{1}_{-1} \, , 
\qquad \bf{8}_{\rm s} = \bf{4}_{-1/2} \oplus \bar{\bf{4}}_{+1/2} \, ,
\]
corresponding to the six scalars, the two helicities of the gluons and
the two helicities of the four gauginos.}. 

In the superspace description of the theory we assign U(1) charge to
the fermionic coordinates -- the $\theta^m$'s have charge $+1/2$ and
the $\bth_m$'s have charge $-1/2$. As a result the superfield
$\phi(x,\th,\bth)$ has definite helicity $+1$, as shown by the
component expansion (\ref{N4superfield}). Similarly, the expression of
the conjugate superfield, $\bar\phi(x,\th,\bth)$, in terms of
component fields shows that it has helicity $-1$.  Notice also that
each $\del$ derivative (and each $p$ component of momentum)
contributes one unit of U(1) charge; each $\bar\del$ derivative (and
each $\bar p$ component of momentum) carries U(1) charge $-1$. The
$\del_\pm$ derivatives and the $p_\pm$ components of momentum do not
carry any U(1) charge. This ensures that the $\calN=4$ SYM action
(\ref{n=4}) is U(1) neutral as required by Lorentz invariance. 

In view of these helicity assignments for the $\calN=4$ superfields,
$\phi$ and $\bar\phi$, we can write the light-front Lagrangian as
\be
L_{\phi,\bar\phi} = \int_\S \dr^3x \,\dr^4\th\,\dr^4\bth \, \left[ 
\calL^{(-+)}_{\phi,\bar\phi} + \calL^{(-++)}_{\phi,\bar\phi} + 
\calL^{(--+)}_{\phi,\bar\phi} + \calL^{(--++)}_{\phi,\bar\phi} \right] \, ,
\label{Lagr-helicities}
\ee
where the integration is on a surface of constant $x^+$ and the
superscripts refer to the number of superfields of helicity $+1$
($\phi$) and $-1$ ($\bar\phi$). Comparing to (\ref{Lagradensity}) we
find
\be
\calL^{(-+)}_{\phi,\bar\phi} = -2\,\Tr\!\left(\bar\phi
\frac{\Box}{\del_-^2}\phi\right) \, ,
\ee
\be
\calL^{(-++)}_{\phi,\bar\phi} = 
i\frac{8}{3}g \, \Tr \!\left(
\fr{\del_-}\bar\phi[\phi,\bar\del\phi] \right) \, , \quad 
\calL^{(--+)}_{\phi,\bar\phi} =  
i\frac{8}{3}g \, \Tr \!\left(
\fr{\del_-}\phi[\bar\phi,\del\bar\phi] \right) 
\ee
and
\be
\calL^{(--++)}_{\phi,\bar\phi} = 2 g^2\,\Tr\!\left(\fr{\del_-}
[\phi,\del_- \phi] \fr{\del_-}[\bar \phi,\del_-\bar\phi] + \half
[\phi,\bar\phi][\phi,\bar\phi] \right) \, .
\ee

\section{Towards a MHV Lagrangian for $\mathcal N=4$ SYM: superfield
redefinition}
\label{SFredef}

In this section, we identify a superfield redefinition
\be
\phi(x,\th,\bth) \to \chi(x,\th,\bth) \, , \qquad
\bar\phi(x,\th,\bth) \to \wtilde\chi(x,\th,\bth) \, ,
\label{fieldredef}
\ee
such that in terms of the new superfields the $\calN=4$ action takes a
manifestly MHV form. We require that the redefinition be a canonical
transformation in superspace. This will ensure that the change of
variables (\ref{fieldredef}) does not give rise to a Jacobian when used in
the path integral. It will also be important that the transformation
preserve the helicity of the superfields, so that $\chi(x,\th,\bth)$
and $\wtilde\chi(x,\th,\bth)$ have the same definite helicities, $+1$
and $-1$ respectively, as the original superfields.  

Our construction of the superfield redefinition (\ref{fieldredef})
follows closely that of \cite{PM,EM} for the pure Yang--Mills case. As
in those papers, we will find that, in order to produce a manifestly
MHV Lagrangian, the redefinition (\ref{fieldredef}) is necessarily non
polynomial. The superfields $\phi$ and $\bar\phi$ are given by
infinite series in the new fields $\chi$ and $\wtilde\chi$. We will
show that the superfield redefinitions take the form~\footnote{Here
and in the following $\dr^3p$ denotes $\dr p_-\,\dr p\,\dr\bar p$ in
momentum space integrals.} 
\ba
&&\hsp{-0.7} \phi(p) = \sum_{n=2}^\infty g^{n-2}\!\!\int\! \dr^3p_1
\cdots\dr^3p_{n-1}\,\delta^{(3)}(p-p_1-\cdots -p_{n-1})
\Gamma(p;p_1,\ldots,p_{n-1})\,\chi(p_1)\cdots\chi(p_{n-1})  \nn \\
&&\hsp{-0.7}\bar\phi(-p) = -\sum_{n=2}^\infty\sum_{s=2}^n g^{n-2} 
\int\dr^3p_1\cdots\dr^3p_{n-1}\,\d^{(3)}(p+p_1+\cdots+p_{n-1}) 
\frac{p_-}{(p_s)_-} \nn \\
&& \hsp{3.7} \times\,\Xi^{(s-1)}(p;p_1,\ldots,p_{n-1}) \chi(p_1)\cdots
\chi(p_{s-1})\wtilde\chi(p_s)\chi(p_{s+1})\cdots\chi(p_{n-1}) \nn 
\ea
where the dependence on the fermionic coordinates, $\th$ and $\bth$,
has not been indicated explicitly. We will derive the explicit form of
the coefficient functions, $\Gamma$ and $\Xi$, in these series using
recursion relations. 

Substituting the expressions of $\phi$ and $\bar\phi$ in terms of
$\chi$ and $\wtilde\chi$ gives rise to a Lagrangian of the form
\be
L_{\chi,\wtilde\chi} = \int_\S \dr^3x\,\dr^4\th\,\dr^4\bth \left[
\calL^{(-+)}_{\chi,\wtilde\chi} + \sum_{k=1}^\infty 
\calL_{\chi,\wtilde\chi}^{(--\overbrace{+\cdots+}^k)} \right]
\ee
in which all terms are manifestly MHV in light-cone superspace. Here
the $k$-th term in sum contains two $\wtilde\chi$'s, $k$ $\chi$'s and
a factor of $g^k$. 

Our superspace analysis presents additional complications, which do
not arise in the non-supersymmetric case. In order to obtain an action
which manifestly displays both the MHV structure and the full
$\calN=4$ supersymmetry, we work with $\phi$ and $\bar\phi$ -- without
eliminating the latter via the constraints (\ref{origconstraints}) --
and construct the map (\ref{fieldredef}) expressing them in terms of
$\chi$ and $\wtilde\chi$. These superfields, however, do not satisfy
the same constraints (\ref{origconstraints}) and so they are not
guaranteed to describe the same degrees of freedom. We therefore need
to identify new constraints satisfied by the transformed superfields
and prove that $\chi$ and $\wtilde\chi$ with these new constraints
describe the irreducible $\calN=4$ multiplet.  This is done in section
\ref{sec:constraints} with further details in appendix
\ref{more-on-constraints}. In section \ref{components} we then show
that the MHV Lagrangian written in terms of $\chi$ and $\wtilde\chi$
reproduces the known terms when expanded in component fields.

\subsection{Canonical Transformation}
\label{sec:phi}

As in the pure Yang--Mills case~\cite{GR,PM,EM}, the aim is to
construct the  superfield redefinition in such a way as to eliminate
the non-MHV cubic vertex, $\calL^{(-++)}_{\phi,\bar\phi}$, from the
Lagrangian. The new superfields, $\chi$ and $\wtilde\chi$, are thus
defined requiring
\be 
\calL^{(-+)}_{\phi,\bar\phi} + \calL^{(-++)}_{\phi,\bar\phi} \to
\calL_{\chi,\wtilde\chi}^{(-+)} \, .
\label{chidef}
\ee
As a preliminary step we re-write the non-MHV vertex,
$\calL^{(-++)}_{\phi,\bar\phi}$, in an equivalent form using
(\ref{new3vertex}). Then the condition (\ref{chidef}) takes the
explicit form~\footnote{From now on we will omit the subscript $\S$
indicating that the integrals are performed on a surface of constant
$x^+$.} 
\be
\label{map}
\int \dr^3x\,\dr^4\theta\,\dr^4\bar\theta\, \Tr \!\left(\! -2\bar \phi 
\frac{\square}{\parm^2} \phi + i\frac{8}{3}g \fr{\parm}\bar\phi 
\fr{\del_-} [\del_- \phi ,\bar\partial \phi]\! \right) \!=\! 
\int  \dr^3x\,\dr^4\theta\,\dr^4\bar\theta \,\Tr \!
\left(\!-2\widetilde \chi \frac{\square}{\parm^2} \chi \!\right) .
\ee
To ensure the canonicity of the transformation, we define the new
superfields, $\chi$ and $\wtilde\chi$, via a generating functional.
In complete analogy with the pure Yang--Mills case~\cite{PM}, we
search for a generating functional of the form
\be
G(\chi,\pi_\phi)=\int\dr^3x\,\dr^4\th\,\dr^4\bth\;\Tr\left[ 
g(\chi)\,\pi_\phi\right] \,, 
\label{genfunctional}
\ee
where $\pi_\phi$ is the
conjugate momentum to $\phi$. From $G(\chi,\pi_\phi)$ we obtain
\be
\label{canpos}
\fr{\parm}\widetilde\chi(x,\theta,\bar\theta\,) \!=\!\! \int\!\! 
\dr^3x^\prime \dr^4\alpha\,\dr^4\bar\alpha\, \frac{\delta \left\{
\phi^a(x^\prime,\alpha,\bar\alpha)(t^a)^i{}_j\!\right\} }{\delta 
\chi(x,\theta,\bar\theta\,)} \fr{\parm}\!\!\left\{
\bar \phi^b(x^\prime,\alpha,\bar\alpha)(t^b)^j{}_i\right\}.
\ee
The form of the generating functional (\ref{genfunctional}) ensures that 
\be
\int\dr^3x\,\dr^4\th\,\dr^4\bth\, \Tr\left(-2\bar\phi\,
\frac{\del_+\del_-}{\del_-^2}\,\phi\right) = 
\int\dr^3x\,\dr^4\th\,\dr^4\bth\, \Tr\left(-2\wtilde\chi\,
\frac{\del_+\del_-}{\del_-^2}\,\chi\right) \, ,
\ee
or, equivalently, using (\ref{integrbyparts}),
\be
\int\dr^3x\,\dr^4\th\,\dr^4\bth\, \Tr\left(2\fr{\del_-}\bar\phi\,
\del_+\phi \right) = 
\int\dr^3x\,\dr^4\th\,\dr^4\bth\, \Tr\left(2\fr{\del_-}\wtilde\chi\,
\del_+\chi \right) \, .
\label{del+term}
\ee
In the following we will therefore ignore the terms involving
$\del_+$. 

The superfield redefinition is more conveniently written in terms of
Fourier transforms, so from now on we work in momentum space. Notice
that our starting point is the Lagrangian (\ref{Lagr-helicities}),
which is written as an integral over a surface of constant $x^+$. In
going to momentum space, we Fourier transform in the $x^-$, $x$ and
$\bar x$ variables only. In all the following expressions the
superfields  have an additional dependence on the time coordinate,
$x^+$, which will be left implicit (see also appendix
\ref{usefulformulae}). In momentum space the condition (\ref{map})
becomes
\bea
&&\int \dr^3p_1\, \dr^4\theta\, \dr^4\bar\theta\; \fr{p_{1-}} 
\bar\phi(-p_1,\theta,\bar\theta) \left\{ \frac{p_1\bar p_1}{p_{1-}}
\,\phi(p_1,\theta,\bar\theta) \right. \nn \\ 
&&+\left. i\frac{2}{3}g\int \dr^3p_2\, \dr^3p_3\;\delta^{(3)}(p_1-p_2-p_3) 
\frac{\left( \bar p_3 p_{2-} - \bar p_2 p_{3-} \right)}{p_{2-}+p_{3-}}
\: \phi(p_2,\theta,\bar\theta)\; \phi(p_3,\theta,\bar\theta) 
\right\} \nn \\
&&= \int \dr^3p\, \dr^4\alpha\, \dr^4\bar\alpha\; \fr{p_-}\, 
\widetilde \chi(-p,\alpha,\bar\alpha)\; \frac{p\bar p}{p_-} \; 
\chi(p, \alpha,\bar \alpha) \, .
\eea
We simplify the notation by writing this as
\bea
\label{map mom}
\int_{1}\int\!\dr\th \fr{p_{1-}} \bar\phi_{-1}\!\! \left\{ \omega_1\phi_1 + 
i\frac{2}{3}g\int_{23}\!\!\!\delta^{(3)}(p_1-p_2-p_3)\, 
\frac{\{2,\!3\}}{p_{2-}+p_{3-}}\,\phi_2\phi_3 
\right\} \!\!=\!\! \int_p\int\!\dr\th \fr{p_-} \widetilde\chi_{-p}\, 
\omega_p \chi_p,
\eea
where $\phi_{j} \equiv \phi(p_j)$, $\phi_{-j} \equiv \phi(-p_j)$,
$\omega_k \equiv  \frac{p_k\bar p_k}{p_{k-}}$ and $\{i,\!j\} \equiv
\left(\bar p_jp_{i-}-\bar p_i p_ {j-}\right)$ and for the measures we
have defined
\be
\int_{12\ldots n} \equiv  \int \prod_{k=1}^n \dr p_{k-}\,\dr p_k\, 
\dr\bar p_k \qquad {\rm and} \qquad \int\dr\th \equiv
\int \dr^4\th\,\dr^4\bth \, .
\ee
In the following we will also use the notation $(p_i,p_j)=p_j {p_i}_-
- p_i {p_j}_-$.  We now substitute (the Fourier transform of)
(\ref{canpos}) into (\ref{map mom}) to obtain
\bea
\label{phi eqn}
\omega_1\phi_1 + i\frac{2}{3}g\int_{23}\delta^{(3)}(p_1-p_2-p_3)\, 
\frac{\{2,\!3\}}{p_{2-}+p_{3-}}\, \phi_2\phi_3  
\!-\!\! \int_{l} \omega_l \frac{\delta \phi_1}
{\delta \chi_l}\, \chi_l = 0,
\eea
which indicates that $\phi$ is a power-series in $\chi$ of the
following form
\be
\label{phi ansatz}
\phi_1 = \sum_{n=2}^{\infty} g^{n-2} \int_{2\ldots n} 
\delta^{(3)}(p_1-p_2-\cdots-p_n)\; \Gamma_{1,2,\ldots,n}\; 
\chi_2\ldots\chi_n\ .
\ee
$\Gamma_{1,2,\ldots,n} \equiv \Gamma(p_1,p_2,\ldots,p_n)$ are
coefficients to be determined order by order. The various permutations
of the  $\chi$ superfields are accounted for by the structure of the
$\Gamma$  coefficients. 

At lowest order, we see from the
canonical constraint~\cite{EM} that $\phi=\chi$, implying that
$\Gamma_{j,k}\,=\,\delta_{j,k}$.  This may be verified by substituting
$\phi_1 =\chi_1$ in (\ref{phi eqn}) at  order $g^0$. We now move to
order $g$ and substitute (\ref{phi ansatz}) in  (\ref{phi eqn}) to
obtain
\be
\Gamma_{p,q,r} = -i \frac{2}{3} \fr{\left(\omega_p -\omega_q -\omega_r
\right)}\frac{\{q,\!r\}}{(q_-+r_-)}\ .
\ee
If we use conservation of momentum implied by the delta function, 
the $\Gamma$ coefficient simplifies to
\be
\Gamma_{p,q,r} = i \frac{2}{3} \frac{q_- r_-}{(q,\!r)},
\label{Gamma3}
\ee
where $\bm{p}=\bm{q}+\bm{r}$. Here and
in the following we use bold-face symbols for the three ``non $+$''
components of vectors, \eg $\bfp=(p_-,p,\bar p)$. 

Then, the field redefinition to order $g$ reads
\be
\label{phi}
\phi_{p} = \chi_{p} 
+ i \frac{2}{3}g\int \dr^3q\, \dr^3r\; \delta^{(3)}(p-q-r)\, 
\frac{q_- r_-}{(q,\!r)}\, \chi_{q}\, \chi_{r}\, .
\ee
An all-order result for the field redefinition is
straightforward to  derive. We substitute (\ref{phi ansatz}) into
(\ref{phi eqn}), and relabel  momentum variables, to obtain the
following recurrence relation for the $ \Gamma$ coefficients in
equation (\ref {phi ansatz})
\be
\Gamma_{1,2,\ldots,n} = -\frac{2i}{3 \,\omega}
\sum_{j=2}^{n-1} \frac{\left\{ \left(p_2+\cdots+p_j\right),
\left( p_{j+1}+\cdots+p_n\right) \right\}}{p_{2-}+p_{3-}+
\cdots+p_{n-}}\:\Gamma_{+,2,\ldots,j}\:
\Gamma_{+,{j+1},\ldots,n} \, , \quad n\ge 3 \, ,
\label{recurrence}
\ee 
where $\Gamma_{+,i,j}\,\equiv\,\Gamma_{i+j,i,j}$, and
$\omega \equiv {\omega_1-\omega_2-\cdots-\omega_n}$. As is to be
expected, there  are many manifest similarities with the all-order
result for the field redefinition  of the gauge field in pure
Yang--Mills theory~\cite{PM}.

Using the momentum conserving delta functions helps simplify the 
expressions. We get
\bea
\Gamma_{+,2,3} =  i \frac{2}{3} \frac{p_{2-} p_{3-}}{(p_2,\! p_3)} \, ,
\qquad  \Gamma_{+,2,3,4} = \left(i \frac{2}{3}\right)^2  
\frac{p_{2-} p_{3-}^2 p_{4-}}{(p_2,\!p_3)(p_3,\!p_4)} \, .
\label{first gammas}
\eea
The general coefficient in (\ref{phi ansatz}) is 
\be
\Gamma_{+,2,\ldots,n} = \left(i \frac{2}{3}\right)^{n-2} 
\frac{ p_{2-} p_{3-}^2 p_{4-}^2 \ldots p_{(n-1)-}^2 
p_{n-}}{(p_2,\!p_3)(p_3,\!p_4)\ldots(p_{n-1},\!p_n)}\, .
\label{genGamma}
\ee
We prove this by induction in appendix \ref{app:Gamma}.  

Having obtained an all-order expression for the field
redefinition for $\phi$ we now turn to $\bar\phi$. We differentiate
$\phi$ with respect to $\chi$ and substitute the result in
(\ref{canpos}) to obtain the following expression for $\bar \phi$
\be
\label{phi bar ansatz}
\bar \phi_{-1} = -\!\sum_{m=2}^{\infty}g^{m-2}
\sum_{s=2}^m \int_{2\ldots m}\!
\delta^{(3)}(p_1+p_2+\cdots+p_m) \frac{p_{1-}}{p_{s-}} \,
\Xi^{s-1}_{1,2,\ldots,m} \chi_{2}\ldots\chi_{s-1} 
\widetilde\chi_{s} \chi_{s+1} \ldots \chi_{m},
\ee
where the superscript on $\Xi$ corresponds to the position of
$\widetilde \chi $ in the string of $\chi$'s. For example, the
coefficient $\Xi^3_{12\ldots n}$  accompanies the string
$\chi_2\chi_3\widetilde\chi_4\chi_5\ldots \chi_n$. Note that
$\Xi_{j,k}^1 \equiv \delta_{j,-k}$. To compute the higher order $\Xi$
coefficients, we start with (\ref{del+term}). From the expansion of
$\phi$ in (\ref{phi ansatz}), since all the  fields have the same
$x^+$ dependence and none of the $\Gamma$ coefficients depend on
$x^+$, we get 
\be
\label{delplus phi1}
\parp \phi_1 = \sum_{n=2}^{\infty} g^{n-2}\sum_{r=2}^n 
\int_{2\ldots n}\!\!\!\!\!\! \delta^{(3)}(p_1-p_2-\cdots-p_n)\, 
\Gamma_{1,2,\ldots,n}\chi_2\ldots \chi_
{r-1}\parp\chi_r \chi_{r+1} \ldots \chi_n.
\ee
We substitute (\ref{phi bar ansatz}) and (\ref{delplus phi1}) in (\ref
{del+term}), and verify that the equation is trivially satisfied at
zero-th order.  Evaluating at order $g$, we find
\bea
\Xi^1_{1,2,3} = -\Gamma_{-2,3,1}\,, \qquad
\Xi^2_{1,2,3} = -\Gamma_{-3,1,2}\,,
\eea
so that
\be
\label{phi bar}
\bar \phi_{-p} = \widetilde \chi_{-p} + g\int \dr^3q\,\dr^3r\, 
\delta^{(3)}(p+q+r) \left\{ \frac{p_-}{q_-}\, 
\Gamma_{-q,r,p}\;\;\widetilde\chi_q\chi_r + \frac{p_-}{r_-}\,
\Gamma_{-r,p,q}\;\; \chi_q\widetilde\chi_r \right\} 
+ \mathcal{O}(g^2) \, .
\ee
It is possible, though tedious, to write a general expression for
$\displaystyle{\Xi^{s-1}_{12\ldots n}}$. We present the details  in
appendix \ref{phibar}.

The transformations (\ref{phi ansatz}) and (\ref{phi bar ansatz}) can
be inverted to express $\chi$ and $\wtilde\chi$ in terms of $\phi$ and
$\bar\phi$. These inverse relations schematically take the form
\be 
\chi(p) = \sum_{n=2}^\infty g^{n-2}\!\!\int\! \dr^3p_1
\cdots\dr^3p_{n-1}\,\delta^{(3)}(p-p_1-\cdots -p_{n-1})
\Delta(p;p_1,\ldots,p_{n-1})\,\phi(p_1)\cdots\phi(p_{n-1}) 
\label{invers1}
\ee
\ba
&&\hsp{-0.6}\wtilde\chi(p) = \sum_{n=2}^\infty\sum_{s=2}^n g^{n-2} 
\int\dr^3p_1\cdots\dr^3p_{n-1}\,\d^{(3)}(p+p_1+\cdots+p_{n-1}) 
\label{invers2} \\
&& \hsp{3.7} \times\,\Upsilon^{(s-1)}(p;p_1,\ldots,p_{n-1}) 
\phi(p_1)\cdots\phi(p_{s-1})\bar\phi(p_s)\phi(p_{s+1})
\cdots\phi(p_{n-1})\,, \nn 
\ea
\ie $\chi$ is a power series in $\phi$ only and $\wtilde\chi$ is a
power series in $\phi$ and $\bar\phi$, with each term containing one
factor of $\bar\phi$. The coefficients, $\Delta$ and $\Upsilon$, in
these equations can be expressed in terms of the coefficients $\Gamma$
and $\Xi$ that we previously determined.

\subsection{New constraints}
\label{sec:constraints}

As discussed in section \ref{N4LCsuperspace} and appendix
\ref{more-on-constraints} the original superfields, $\phi$ and
$\bar\phi$, are both constrained. Explicitly, they satisfy (anti)
chirality conditions,
\be
\label{con1}
d^m\,\phi=0\ , \qquad \bar d_m\,\bar\phi=0\ ,
\ee
and what is referred to as an ``inside-out'' relation,
\be
\bar d_m\,\bar d_n\, \bar d_p\,\bar d_q\, \phi = 
2\, \veps_{mnpq}\, \parm^2\, \bar \phi\,.
\label{con2}
\ee
In fact, one verifies that (\ref{con2}), together with the
supersymmetry algebra (\ref{dsusyalg}), gives rise to the additional
``hidden'' constraints
\bea
\label{con3}
\bar d_m\, \bar d_n\, \bar d_p\, \phi &=& i\sqrt{2}\, \veps_{mnpq}\, d^q\, 
\parm\, \bar \phi\;, \\
\label{con4}
\bar d_m\, \bar d_n\, \phi &=& \fr{2}\, \veps_{mnpq}\, d^p\, d^q\, 
\bar\phi\;, \\
\label{con5}
\bar d_m\, \phi &=& \frac{i}{6\sqrt{2}}\, \veps_{mnpq}\,
d^n\,d^p\,d^q\, \fr{\parm}\, \bar \phi \\
\label{con6}
\phi &=& \fr{96} \, \veps_{mnpq} d^m \, d^n \, d^p \, d^q \, 
\fr{\del_-^2} \,\bar\phi \, .
\eea
The most general superfield in $\mathcal N=4$ superspace does not
describe an irreducible multiplet of the $\calN=4$
superalgebra. Imposing the constraints (\ref{con1})-(\ref{con6})
reduces the number of independent components in $\phi$ and $\bar\phi$
ensuring that these superfields describe {\it  only} the $\mathcal
N=4$ degrees of freedom. 

In the previous subsection we have constructed the superfield
redefinition (\ref{fieldredef}) to all orders and in the next section
we will present the manifestly MHV $\calN=4$ Lagrangian written in
terms of the new superfields, $\chi$ and $\wtilde\chi$.  However, we
first need to show that these new superfields also describe the
$\calN=4$ supermultiplet. This is not guaranteed, because in
constructing the canonical change of variables, we have treated $\phi$
and $\bar\phi$ as unconstrained. We need to deduce what conditions for
$\chi$ and $\wtilde\chi$ are implied by the constraints on the
original superfields and then show that these new conditions give rise
to the correct degrees of freedom. This can be achieved starting with
the inverse transformations (\ref{invers1})-(\ref{invers2}) and
imposing the conditions (\ref{con1})-(\ref{con6}) on the right hand
side.  

From the transformation relating $\phi$ and $\chi$ one can verify that
the latter is also chiral,
\be
\label{0}
d\, \chi = 0 \, .
\ee
This is shown to be valid to all orders in $g$ in appendix
\ref{constraint}. 
 
The remaining constraints on $\phi$ and $\bar\phi$ are, however, not
valid for $\chi$ and $\wtilde\chi$. In particular, the superfield
$\wtilde\chi$ is not anti-chiral. Moreover, as a consequence of the
structure of the field redefinition, we expect the constraints
satisfied by $\chi$ and $\wtilde\chi$ to be modified order by order in
the coupling. We will present here the schematic form of the new
conditions for $\chi$ and $\wtilde\chi$ to order $g$. Appendix
\ref{more-on-constraints} contains more details of the derivation. 

We start with the inverse transformations
(\ref{invers1})-(\ref{invers2}) truncated at order $g$,
\bea
\chi_p &=& \phi_p - g\int_{qr} \delta(p-q-r)\; \Gamma_{p,q,r}\;
\phi_q\, \phi_r + \calO (g^2) \;,  
\label{invredef1} \\
\widetilde \chi_p &=& \bar \phi_p + g\int_{qr}\!\! \delta(p-q-r) 
\left\{ \frac{p_-}{q_-}\; \Gamma_{-q,r,-p}\; \bar \phi_q\, \phi_r\;
+ \frac{p_-}{r_-}\; \Gamma_{-r,-p,q}\; \phi_q \, \bar \phi_r \right\} 
+ \calO (g^2)\,.
\label{invredef2}
\eea
The expansion (\ref{invredef1}) is of course consistent with the
chirality of $\chi$. Acting with the superspace derivative ${\bar
d}_m$ on (\ref{invredef2}) and using (\ref{con1})-(\ref{con6}) we
arrive at the relation (see appendix \ref{constraint} for further
details)
\be
\label{dbar chit}
\fr{\del_-^2}{\bar d} \,{\wtilde\chi} \sim g \left(
\frac{d^3}{\del_-}\wtilde\chi\right)\wtilde\chi \, ,
\ee
which replaces the anti-chirality condition for $\bar\phi$. 

The additional constraint relations, analogous to
(\ref{con2})-(\ref{con6}), are
\bea
\label{io1}
\frac{d^4}{\parm^2}\widetilde \chi &\sim& \chi + g\chi^2 + g^3\chi^3 + 
\cdots\,, \\
\label{io2}
\frac{d^3}{\parm}\widetilde \chi &\sim& \bar d \chi + g\chi 
\bar d \chi + g^2 \chi^2\bar d \chi+\cdots\,, \rule{0pt}{12pt} \\
\label{io3}
d^2\widetilde \chi &\sim& \bar d\,^2\chi + g\bar d\chi\bar d \chi + 
g\chi\bar d\,^2\chi + g^2\chi\bar d \chi\bar d \chi + g^2\chi^2
\bar d\,^2\chi+\cdots\,, \rule{0pt}{12pt} \\
\label{io4}
\parm d\widetilde \chi &\sim& \bar d\,^3\chi + g\bar d\chi 
\bar d\,^2\chi + g\chi\bar d\,^3 \chi + \cdots\,, \rule{0pt}{16pt} \\
\label{io5}
\parm^2 \widetilde \chi &\sim& \bar d\,^4 \chi + g\bar d\,^2\chi
\bar d\,^2\chi + g\bar d\chi\bar d\,^3\chi + g\chi\bar d\,^4\chi 
+ \cdots\,. \rule{0pt}{16pt}
\eea 
These are derived in appendix \ref{insideout}.

Notice that at zero-th order in the coupling $\chi$ and $\wtilde\chi$
coincide with $\phi$ and $\bar\phi$ respectively. The above conditions
are consistent with this observation. The superfield $\chi$ is chiral
and (\ref{dbar chit}) reduces to
\be
{\bar d} \,(\wtilde\chi)_0 = 0 \, ,
\ee
showing that $\wtilde\chi$ is anti-chiral for $g=0$. Similarly the
conditions (\ref{io1})-(\ref{io5}) reduce to
(\ref{con2})-(\ref{con6}) at $g=0$.

Having derived the new constraints satisfied by $\chi$ and
$\wtilde\chi$ we proceed to show that they give rise to the correct
field content. Since $\chi$ is chiral, we can write it
as~\footnote{Here and in the following we use the notation $[\th]^n$
to denote $n$ powers of $\th$ without specifying the SU(4)$_R$
indices.}
\be
\label{chiralchi}
\chi(x,\th,\bth) = \alpha(y) + \beta(y) \,[\th] + \gamma(y) \,[\th]^2
+ \delta(y) \,[\th]^3 + \veps(y) \,[\th]^4\,.
\ee
We find that $\widetilde\chi$ satisfying  the ``inside-out'' relations 
(\ref{io1})-(\ref{io5}) is forced to have the structure
\bea
\label{chitildestr1}
\widetilde\chi(x,\th,\bth) &\!\!=\!\!& A_{00}(y) + A_{10}(y) \,[\th]
+ A_{20}(y) \,[\th]^2 + A_{30}(y) \,[\th]^3 + A_{40}(y) \,[\th]^4 \nn \\
&&+ A_{01}(y) \,[\bth] + A_{02}(y) \,[\bth]^2 + A_{03}(y) \,[\bth]^3 + 
A_{04} (y) \,[\bth]^4 \, , \rule{0pt}{15pt}
\eea
where $y$ is the chiral variable (\ref{chirvar}) and all component
fields, $A_{ij}$, $i,j=0,\ldots,4$, are fully determined in terms of
the component fields  $\alpha,\beta,\gamma,\delta$ and $\veps$. 

The remaining condition on $\chi$ and $\wtilde\chi$ is (\ref{dbar
chit}). Imposing this constraint halves the number of independent
components in the new superfields. Therefore we conclude that $\chi$
and $\wtilde\chi$ contain a total of eight bosonic and eight fermionic
independent degrees of freedom.

\section{MHV Lagrangian for $\mathcal{N}=4$ Yang--Mills}
\label{sec:MHVlag}

The manifestly MHV Lagrangian in terms of the new superfields 
$\chi$ and $\widetilde\chi$ to  order $g^2$ is
\bea
L_{\chi,\widetilde\chi} &=&\Tr\int \dr\theta\, \dr p\; f_0(p)\; 
\widetilde\chi_{-p}\, \chi_p \nn \\ \nn \\
&&+ g\;\Tr \int \dr\theta\,\dr p\,\dr k\,\dr l\,  \delta(p+k+l)\; 
f_1(p,k,l)\; \chi_p\,\widetilde\chi_k\,\widetilde\chi_l  \nn \\ \nn \\
&&+ g^2\, \Tr \int \dr\theta\,\dr p\,\dr q\,\dr r\,\dr l\,
\delta(p+q+r+l)\; f_{21}(p,q,r,l) \; \chi_p\,\chi_q\,
\widetilde\chi_r\,\widetilde\chi_l \nn \\
&&+ g^2\, \Tr\int \dr\theta\,\dr p\,\dr q\,\dr r\,\dr l\,
\delta(p+q+r+l)\; f_{22}(p,q,r,l) \; \chi_p\,\widetilde\chi_q\,
\chi_r\,\widetilde\chi_l\,, \nn \\ 
\eea
where 
\bea
f_0(p) &=&  -4\;\frac{p\bar p-p_+p_-\,}{p_-^2}\,,\\ \nn \\
\label{eq:f1}
f_1(p,q,r) &=&  \frac{i8}{3}\;\frac{(q,r)}{p_-(q_-+r_-)}\,, \\ \nn \\
f_{21}(p,q,r,l) &=&  f_1(p+q,r,l)\,\Gamma(p+q,p,q) - f_1(p,q+r,l)
\left(\frac{q_-+r_-}{r_-}\right)\Gamma(-r,-q-r,q)  \nn \\ 
&& - f_1(q,r,p+l)\left(\frac{l_-+p_-}{l_-}\right)\Gamma(-l,p,-l-p) \nn \\
&&  +2 \left[\left(\frac{q_--p_-}{q_-+p_-}\right)\left(
\frac{l_--r_-}{l_-+r_-}\right)-1\right]\,, \\ \nn \\
f_{22}(p,q,r,l) &=& -f_1(p,q,r+l)\left(\frac{l_-+r_-}{l_-}
\right)\Gamma(-l,-l-r,r) +2 \nn \\
&&- f_1(p,q+r,l)\left(\frac{q_-+r_-}{q_-}\right)\Gamma(-q,r,-q-r)\,.
\eea
We now substitute the expressions for the $\Gamma$
coefficients. Employing the delta  function and writing $f_{21}$ and
$f_{22}$ in terms of the independent  momentum variables, we obtain
\ba
\label{eq:f21}
f_{21}(p,q,r,l) &\!\!=\!\!& \frac{16}{9}
\frac{p_-(q_- + r_-)^2 (q,r)}{q_-^2 l_- (q+r,l)}
+\frac{16}{9}\frac{q_- (q_-+r_-)^2 (q+r,l)}{ p_-^2r_- (q,r)} \nn \\ 
&& -\frac{16}{9} \frac{q_-p_- (r,l)}{(r_-+l_-)^2 (r+l,q)} + 
\frac{4 (q_-r_-+l_-p_-)}{(r_-+l_-)^2} \,, \\ 
\label{eq:f22}
f_{22}(p,q,r,l)&=& 2+\frac{16}{9}\frac{r_-}{p_-^2}\left[
\frac{(q_-+r_-)^2 (q+r,l)}{q_- (q,r)}+\frac{(l_- + r_-)^2 
(l+r,q)}{l_-(l,r)}\right]\,.
\rule{0pt}{25pt}
\ea
In order to make the colour structure of the new Lagrangian more
transparent, it is convenient to write the interaction vertices in
terms of commutators,
\bea
L_{\chi,\widetilde\chi} &=&\Tr\int \dr\theta\, \dr p\; f_0(p)\; 
\widetilde\chi_{-p}\, \chi_p \nn \\ \nn \\
&&+ g\;\Tr \int \dr\theta\,\dr p\,\dr k\,\dr l\,  \delta(p+k+l)\; 
g_1(p,k,l)\; \chi_p \left[ \widetilde\chi_k\,,\widetilde\chi_l \right]  
\nn \\ \nn \\
&&+ g^2\, \Tr \int \dr\theta\,\dr p\,\dr q\,\dr r\,\dr l\,
\delta(p+q+r+l)\; g_{21}(p,q,r,l) \; \left[ \chi_p\,,\chi_q \right]
\left[ \widetilde\chi_r\,,\widetilde\chi_l \right] \nn \\
&&+ g^2\, \Tr\int \dr\theta\,\dr p\,\dr q\,\dr r\,\dr l\,
\delta(p+q+r+l)\; g_{22}(p,q,r,l) \; \left[ \chi_p\,,\widetilde\chi_q 
\right]\left[ \chi_r\,,\widetilde\chi_l \right]\,.
\label{finalLagrangian}
\eea
Expanding the commutators, and relabelling momenta, it is easy to 
arrive at the following relations
\bea
f_1(p,q,r) &=& g_1(p,q,r)-g_1(p,r,q)\,, \\
f_{21}(p,q,r,l) &=& g_{21}(p,q,r,l)-g_{21}(p,q,l,r)-
g_{21}(q,p,r,l)+g_{21}(q,p,l,r) \nn \\
   && -\, g_{22}(q,r,p,l) - g_{22}(p,l,q,r)\,, \\
f_{22}(p,q,r,l) &=& g_{22}(p,q,r,l) + g_{22}(p,l,r,q)\,.
\eea
From the above the relations, we derive the symmetry properties
of $f_1$, $f_{21}$ and $f_{22}$, as
\bea
f_1(p,q,r) &=& - f_1(p,r,q)\,, \\
f_{21}(p,q,r,l) &=& f_{21}(q,p,l,r)\,, \\
f_{22}(p,q,r,l) &=& f_{22}(p,l,r,q)\,,
\eea
which can be readily verified using the explicit expressions for 
$f_1$, $f_{21}$ and $f_{22}$ given in (\ref{eq:f1}),
(\ref{eq:f21}) and (\ref{eq:f22}).

Our task is now to determine the form of the coefficients 
$g_1$, $g_{21}$ and $g_{22}$. It is easy to see that
\be
\label{eq:g1}
g_1(p,q,r) = \frac{1}{2}\, f_1(p,q,r) =
\frac{i4}{3} \frac{(q,r)}{p_-(q_-+r_-)}\, .
\ee
The derivation of the other two coefficients, $g_{21}$ and $g_{22}$, 
is slightly more lengthy, but straightforward. They are
\bea
g_{21}(p,q,r,l) &=& \frac{16}{9} \frac{p_-}{q_-^2} 
\frac{(q_- + r_-)^2 (q,r)}{l_- (q+r,l)}
-\frac{4}{9} \frac{q_- p_- (r,l)}{(r_- + l_-)^2 (r+l,q)} \,, \\
g_{22}(p,q,r,l) &=& \frac{16}{9} \frac{r_-}{p_-^2} 
\frac{(q_- + r_-)^2 (q+r,l)}{q_- (q,r)} - 2\, 
\frac{(q_-p_- + l_- r_-)}{(q_-+l_-)^2}\,,
\eea
where $\bfp + \bfq + \bfr +  \bfl = 0$.

Using the explicit expressions for the coefficients $\Gamma$ and $\Xi$
in (\ref{phi ansatz}) and (\ref{phi bar ansatz}) it is possible,
though tedious, to derive higher order vertices in the MHV
Lagrangian. We will not present these calculations here.

\section{Component Lagrangian}
\label{components}

In this section we discuss the form of the gluon MHV vertices arising
from the component expansion of the superspace Lagrangian given in the
previous section. These gluon vertices should coincide with those in
the pure Yang--Mills MHV Lagrangian~\cite{PM}. This will thus allow us
to test our superspace result. We will carry out the comparison for
terms up to order $g^2$, \ie we will consider cubic and quartic
vertices. 

We obtained the superfield redefinition requiring that the
transformation be canonical and eliminate the non-MHV cubic vertex, 
\be 
\calL^{(-+)}_{\phi,\bar\phi} + \calL^{(-++)}_{\phi,\bar\phi} \to
\calL_{\chi,\wtilde\chi}^{(-+)} \, . \nn
\ee
The explicit form of the redefinition derived in section \ref{sec:phi}
to order $g$ is
\bea
\label{redef}
\phi_p &=& \chi_p + g\int_{qr} \delta(p-q-r)\; \Gamma_{p,q,r}\; 
\chi_q\, \chi_r + \calO (g^2) \;, \nn \\
\bar \phi_p &=& \widetilde \chi_p - g\int_{qr} \delta(p-q-r) 
\left\{ \frac{p_-}{q_-}\; \Gamma_{-q,r,-p}\; \widetilde\chi_q\, \chi_r\;
+ \frac{p_-}{r_-}\; \Gamma_{-r,-p,q}\; \chi_q \, \widetilde \chi_r \right\} 
+ \calO (g^2)\,. \nn
\eea
These can easily be inverted to obtain
\bea
\chi_p &=& \phi_p - g\int_{qr} \delta(p-q-r)\; \Gamma_{p,q,r}\;
\phi_q\, \phi_r + \calO (g^2) \;,  \nn \\
\widetilde \chi_p &=& \bar \phi_p + g\int_{qr}\!\! \delta(p-q-r) 
\left\{ \frac{p_-}{q_-}\; \Gamma_{-q,r,-p}\; \bar \phi_q\, \phi_r\;
+ \frac{p_-}{r_-}\; \Gamma_{-r,-p,q}\; \phi_q \, \bar \phi_r \right\} 
+ \calO (g^2)\,.
\label{invredef}
\eea
Since we are focussing on the gluon contributions only, we can set all
other components in $\phi$ and $\bar\phi$ to zero and
use~\footnote{Here and in the following we use the notation $\langle
\theta \rangle^4$ to denote $\veps_{mnpq} \theta^m \theta^n \theta^p
\theta^q$, and similarly for  $\langle \bar \theta \rangle^4$. We will
also write $\left(\theta^m \bar \theta_m\right)$ as just $\left(
\theta \bar\theta \right)$.}
\bea
\phi_p &=& -\frac{1}{ip_-}\; A_p\; \er^{-\frac{i}{\sqrt 2} \ttb i p_-}
- \frac{1}{12} \langle \theta \rangle^4\, ip_-\, \bar A_p \,, \nn \\
\bar \phi_p &=& -\frac{1}{ip_-}\; \bar A_p\; \er^{\frac{i}{\sqrt 2} 
\ttb i p_-}- \frac{1}{12} \langle \bar \theta \,\rangle^4\, 
ip_-\, A_p \,. 
\label{phi-comps}
\eea
In order to make contact with the known form of the MHV gluon
couplings we then need to express the component fields, $A$ and $\bar
A$, in terms of the new fields describing the two helicities of the
gluons, $B$ and $\wtilde B$. We use the form of the field redefinition
derived in~\cite{GR,PM} for the Yang--Mills case. The details of the
calculation are presented in appendix \ref{compdetails}. The form of
$\chi$ and $\wtilde\chi$ in terms of $B$ and $\wtilde B$ is given in
(\ref{chiinB}) and (\ref{chitinB}).  Substituting these expressions
into our superspace Lagrangian (\ref{finalLagrangian}) reproduces
exactly the cubic and quartic vertices in the MHV Lagrangian
of~\cite{GR,PM},
\be
L_{B,\wtilde B}^{(--+)} = -\frac{i}{2}\int_{qr} \delta(p+q+r) \,
\frac{r_-(p,q)^3}{p_-q_-(p,r)(r,q)}
\,\Tr \left(\wtilde B_p\,\wtilde B_q\, B_r \right) 
\ee
and
\ba
&& \hsp{-2} L_{B,\wtilde B}^{(--++)} = \half \int_{qrs} 
\delta(p+q+r+s) \left\{ \frac{2\,r_- s_- (p,q)^4}
{p_- q_- (p,q)(q,r)(r,s)(s,p)}\,\Tr\left(
\wtilde B_p\,\wtilde B_q\,B_r\,B_s \right ) \right. \nn \\
&&\hsp{3.7}\left. + \frac{q_{-}s_{-} (p,r)^4}
{p_{-}r_{-}(p,q)(q,r)(r,s)(s,p)} \, \Tr\left(
\wtilde B_p\,B_q\,\wtilde B_r\,B_s\right) \right\} \, .
\ea

\section{Discussion}
\label{concl}

In this paper we constructed a MHV Lagrangian for $\calN=4$
SYM in light-cone superspace. Through a canonical change of variables
we obtained a non-polynomial Lagrangian consisting of MHV vertices
involving two superfields, denoted by $\wtilde\chi$, of helicity $-1$
and an arbitrary number of superfields of helicity $+1$, denoted by
$\chi$. Our Lagrangian takes the manifestly MHV form
\be
L_{\chi,\wtilde\chi} = \int_\S \dr^3x\,\dr^4\th\,\dr^4\bth \left[
\calL^{(-+)}_{\chi,\wtilde\chi} + \sum_{k=1}^\infty 
\calL_{\chi,\wtilde\chi}^{(--\overbrace{+\cdots+}^k)} \right] 
\label{chichitildLagr}
\ee
and we explicitly determined to all orders the form of the
coefficients in the series expressing the superfields, $\chi$ and
$\wtilde\chi$, in this equation in terms of the original $\calN=4$
superfields, $\phi$ and $\bar\phi$.  Both the derivation and the final
expression we obtained bear a close similarity with the construction
of~\cite{PM,EM} for pure Yang--Mills. In our reformulation of the
theory, as in the original description in light-cone superspace, the
full $\calN=4$ supersymmetry as well as the SU(4)$_R$ R-symmetry are
manifestly realised. 

As mentioned in the introduction, a MHV Lagrangian for $\calN=4$ SYM
in light-cone superspace has previously been proposed
in~\cite{FH}. The approach taken in that paper differs from ours in
some essential respects.  The $\calN=4$ Lagrangian in light-cone
superspace can be written in terms of the single superfield
$\phi$. This is achieved by eliminating $\bar\phi$ using the
constraints (\ref{origconstraints}), which imply
\be
\bar\phi = \fr{48}\frac{\la\bar d\,\ra^4}{\del_-^2}\,\phi \, ,
\ee
with $\la\bar d\,\ra^4 = \veps^{mnpq}{\bar d}_m{\bar d}_n{\bar d}_p{\bar
d}_q$.  This leads to a different form of the $\calN=4$
Lagrangian~\cite{BLN2}, in which $\phi$ has the component expansion
(\ref{N4superfield}), but is otherwise treated as unconstrained. This
form of the Lagrangian was used as a starting point in~\cite{FH}. This
approach has the advantage that one need not worry about constraints
for the single redefined superfield, $\chi$ in our notation. However,
in this formulation the helicity structure of the $\calN=4$ Lagrangian
is somewhat obscured. This is because the helicity content of the
various vertices depends not only on the combination of superfields
they contain, but also on the explicit chiral derivatives, $\bar d_m$,
which carry U(1) charge. We have therefore chosen to take as our
starting point the light-cone superspace Lagrangian
(\ref{n=4})-(\ref{Lagradensity}),  written in terms of both $\phi$ and
$\bar\phi$. As a consequence we had to construct our canonical
transformation on constrained superfields. This led to the need of
determining the new constraints satisfied by the redefined
superfields, $\chi$ and $\wtilde\chi$, in order to ensure that our MHV
Lagrangian describe the correct degrees of freedom. Our approach has,
however, the advantage of making the helicity structure of the new
Lagrangian more transparent, with all vertices being manifestly MHV.

We have shown explicitly up to order $g^2$, that our superspace
Lagrangian reproduces the known gluon vertices in the MHV Lagrangian
for pure Yang--Mills~\cite{GR,PM,EM}. The analysis of the Lagrangian
in component fields, needed for this comparison, revealed an
unexpected  feature. Because the component fields in $\chi$ and
$\wtilde\chi$ are themselves infinite series in the redefined
component fields of~\cite{PM} and their super-partners, we found that
the four gluon MHV vertex gets contribution from both
$\calL^{(--++)}_{\chi,\wtilde\chi}$ and the order $g$ term in
$\calL^{(--+)}_{\chi,\wtilde\chi}$. In general, the MHV vertex with
$n$ positive helicity gluons receives contribution from all the
superspace vertices in (\ref{chichitildLagr}) with $k\le n$. This is a
drawback of our formalism as it limits the usefulness of the
Lagrangian (\ref{chichitildLagr}) as a tool for computing amplitudes. 

However, as stated in the introduction, our main interest is not in
developing a computational tool, but rather in constructing a
formalism which may prove useful in the study of the origin and
implications of the some of the remarkable features of the $\calN=4$
scattering amplitudes. In particular a manifestly supersymmetric
approach appears to be crucial in order to explain the dual
(super)conformal properties of non-MHV amplitudes~\cite{DHKS}. We
therefore believe that our formalism will be useful in the study of
the dual superconformal symmetry. It should also be noted that the
light-cone superspace formulation of $\calN=4$ SYM (and its MHV
version developed in this paper) closely resembles the on-shell
superspace considered in~\cite{DHKS}, while having the advantage of
being suitable for off-shell calculations as well. In view of this we
expect that our results will be useful to further develop the work
initiated in~\cite{corr-Wloop,corr-ampl}.

Another natural application of our formalism is in connection with the
supersymmetry properties the of $\calN=4$ scattering amplitudes
discussed in~\cite{DZFetal}. Both the selection rules for various
types of MHV and N$^k$MHV amplitudes and the supersymmetry and
R-symmetry Ward identities discussed in these papers should have a
simple formulation using our MHV light-cone formalism. 

Scattering amplitudes in theories obtained as marginal deformations of
$\calN=4$ SYM share many of the properties observed in the parent
theory~\cite{VVK}. It will be interesting to study the generalisation
of our results to the case of $\b$-deformations (both supersymmetric
and non) of $\calN=4$ SYM. Such generalisations should be
straightforward using the light-cone superspace formulation of these
theories given in~\cite{AKS}.

\vskip 0.7cm
\ndt 
{\bf {Acknowledgments}}\\[0.3cm]
We thank Lars Brink and Hidehiko Shimada for helpful comments. This
work is supported by the Max Planck Society, Germany, through the Max
Planck Partner Group in Quantum Field Theory. S.A. acknowledges
support by the Department of Science and Technology, Government of
India, through a Ramanujan Fellowship. S.P. is supported by a summer
research fellowship from the Indian Academy of Sciences, Bangalore.

\appendix

\section{Some conventions and useful formulae}
\label{usefulformulae}

We define the Fourier transform of a superfield, $\psi(x,\th,\bth)$,
in light-cone superspace as~\footnote{As usual in computing Fourier
transforms in superspace the fermionic coordinates are left
untouched.}
\be
\hat\psi(x^+;k_-,k,\bar k,\th,\bth) = \int \dr x^-\,\dr x\,\dr\bar x\:
\er^{i\bm{k}\cdot\bm{x}}\,\psi(x^+;x^-,x,\bar x,\th,\bth) \, ,
\label{FourierSF}
\ee
where we indicated explicitly the dependence on the time variable,
$x^+$, which is not transformed and $\bm{k}\cdot\bm{x} =
k_-x^-+kx+\bar k\bar x$.  The inverse Fourier transform is
\be
\psi(x^+;x^-,x,\bar x,\th,\bth) = \int 
\frac{\dr k_-\,\dr k\,\dr\bar k}{(2\pi)^3}
 \: \er^{-i\bm{k}\cdot\bm{x}} \,\hat\psi(x^+;k_-,k,\bar k,\th,\bth) \, .
\label{invFourierSF}
\ee
When working with the action,
\be
\calS = \int \dr x^+\: L_{\phi,\bar\phi} \, ,
\ee
we can further Fourier transform in the time variable. Therefore the
$\calN=4$ action (\ref{n=4}) can be re-written in momentum space as
\ba
&& \hsp{-0.7} \calS = 72\!\int\!\dr^4\th\,\dr^4\bth\,\Tr\left\{ 
\int \frac{\dr^4 p_1}{(2\pi)^4} \, 2 \hat{\bar\phi}(-p_1) 
\frac{p_{1\,\mu}^2}{p_{1\,-}^2} \hat\phi(p_1) \right. \nn \\
&& \hsp{-0.7} - \frac{8ig}{3} \!\int\!\frac{\dr^4p_1}{(2\pi)^4}
\frac{\dr^4p_2}{(2\pi)^4} 
\fr{(p_{1\,-}+p_{2\,-})}\!\left(\bar p_2\,\hat{\bar\phi}(-p_1-p_2)
[\hat\phi(p_1),\hat\phi(p_2)] + p_2
\,\hat\phi(-p_1-p_2)[\hat{\bar\phi}(p_1),
\hat{\bar\phi}(p_2)] \right)  \nn \\
&& \hsp{-0.7} - 2 g^2 \!\int\! \frac{\dr^4p_1}{(2\pi)^4} 
\frac{\dr^4p_2}{(2\pi)^4}\frac{\dr^4p_3}{(2\pi)^4} \left(
\frac{p_{1\,-}p_{3\,-}}{(p_{2\,-}+p_{3\,-})^2} 
[\hat\phi(-p_1-p_2-p_3),\hat\phi(p_1)][\hat{\bar\phi}(p_2),
\hat{\bar\phi}(p_3)] \right. \nn \\ 
&& \hsp{3.85} \left. \left. -\half [\hat\phi(-p_1-p_2-p_3),
\hat{\bar\phi}(p_1)][\hat\phi(p_2),\hat{\bar\phi}(p_3)] \right) 
\right\} \, ,
\ea
where we omitted the $\th$ and $\bth$ arguments in the superfields and 
$p_\mu^2 = 2(-p_+p_-+p\bar p)$.

In superspace integrals the $\fr{\del_-}$ operator can be ``integrated 
by parts''. For generic superfields $f(x,\th,\bth)$ and $g(x,\th,\bth)$ 
we have 
\ba
&&\hsp{-1} \int\dr^{12}z \, f(z)\fr{\del_-}g(z) = 
\int\dr^{12}z\,\frac{\del_-}{\del_-} f(z) \fr{\del_-}g(z) \nn \\
&& \hsp{2.3} = -\int \dr^{12}z\, \fr{\del_-} f(z) \frac{\del_-}{\del_-} g(z) 
= -\int\dr^{12}z\,\fr{\del_-}f(z)g(z) \, ,
\label{integrbyparts}
\ea
where $z=(x^+,x^-,x,\bar x,\th^m,\bth_m)$ and
$\dr^{12}z=\dr^4x\,\dr^4\th\,\dr^4\bth$. 

For any arbitrary function, $X(\phi)$, of the $\calN=4$ superfield,
$\phi$, satisfying (\ref{origconstraints}), the following identity
holds~\cite{abkr}
\be
\int_\S \dr^3x\,\dr^4\th\,\dr^4\bth\: \Tr\!\left(\fr{\del_-^2}
\bar\phi[\phi,X(\phi)] 
\right) = 0 \, .
\ee
Using this relation, the non-MHV cubic vertex in the $\calN=4$ Lagrangian 
can be re-written as
\be
\int_\S \dr^3x\,\dr^4\th\,\dr^4\bth \, \Tr\left(\fr{\del_-}\phi[\bar\phi,
\del\bar\phi] \right) = \int_\S \dr^3x\,\dr^4\th\,\dr^4\bth\, \Tr\left(
\fr{\del_-}\bar\phi\fr{\del_-}[\del_-\phi,\bar\del\phi]\right) \, .
\label{new3vertex}
\ee

\section{Constraints in light-cone superspace}
\label{more-on-constraints}

In this section we discuss the constraint relations used to define
irreducible representations in light-cone superspace. We also
provide further details on the constraints satisfied by the new
superfields, $\chi$ and  $\wtilde\chi$, used in the manifestly MHV 
Lagrangian. 

In section \ref{N4LCsuperspace} we introduced light-cone superspace as
parametrised by the coordinates $(x^+,x^-,x,\bar x,\th^m,\bth_m)$. In
discussing chiral superfields it is convenient to consider the change
of variables
\be
(x^+,x^-,x,\bar x;\theta^m,\bar\theta_m) ~ \to ~ 
(y^+\equiv x^+,y^-\equiv x^--\frac{i}{\sqrt{2}}\th^m\bth_m,y\equiv x,
\bar y\equiv\bar x,\eta^m\equiv\th^m,\bar\eta_m\equiv\bth_m) .
\label{chiralcoord}
\ee
In terms of these `chiral' coordinates the superspace derivatives,
$d^m$ and $\bar d_m$, take the form
\be
d^m = -\frac{\partial}{\partial\bth_m} \, , \qquad
\bar d_m = \frac{\partial}{\partial\th^m} -
i\sqrt{2}\bth_m\del_- \, .
\ee
So a chiral superfield, $\psi(x,\th,\bth)$, as a function of the
coordinates  (\ref{chiralcoord}) satisfies
\be
-\frac{\del}{\del\bth_m} \,\psi(y,\th,\bth) = 0 \, .
\ee
The component expansion of $\psi$, in the original superspace
coordinates, is thus
\be
\psi(x,\th,\bth) = a^{(0)}(y) + a^{(1)}(y)[\theta] + a^{(2)}(y) [\theta]^2
+a^{(3)}(y)[\theta]^3 + a^{(4)}(y) [\theta]^4 = \sum_{n=0}^4 a^{(n)}(y)
[\theta]^n\, ,
\label{chirfield}
\ee
where the right hand side is understood as a power expansion around
$x^-$ and we are using the notation $[\th]^n$ to denote the product of
$n$ $\th$'s.  We can also write $\psi$ as
\be
\psi(x,\th,\bth) = {\rm e}^{-\frac{i}{\sqrt{2}} \theta\bar\theta \partial_-} 
\sum_{n=0}^4 a^{(n)}(x) [\theta]^n \, .
\ee
In momentum space the chiral derivatives (\ref{chiralder}) become
\be
{\hat d}^m_k = -\frac{\partial}{\partial\bar\theta_m} - 
\frac{1}{\sqrt{2}}\theta^m k_- \, , \qquad
{\hat{\bar d}}_{m,k} = \frac{\partial}{\partial\theta^m} 
+ \frac{1}{\sqrt{2}} \bar\theta_m k_-  \, .
\label{ddbarfourier}
\ee
A chiral superfield in momentum space satisfies
\be
{\hat d}^m_k \hat\psi(k,\th,\bth) = 0 
\ee
and has the general expansion
\be
\hat\psi(k,\th,\bth) = \er^{\frac{1}{\sqrt{2}}k_-\th\bth}
\sum_{n=0}^4 {\hat a}^{(n)}(k)[\theta]^n   \, ,
\label{chiralfourier}
\ee
Notice, however, that
\be
{\hat d}^m_k \hat\psi(p,\theta,\bar\theta) \neq 0 \quad {\rm for} \quad
p\neq k \, .
\ee
Therefore products of chiral superfields, such as
\be
\hat\psi(k_1,\th,\bth)\hat\psi(k_2,\th,\bth)\cdots\hat\psi(k_n,\th,\bth) \, ,
\ee
are not chiral. 

\subsection{Chirality of the new superfield}
\label{constraint}

The canonical change of variables which puts the ${\cal N}=4$ action in 
the MHV form gives $\chi$ (in momentum space) as
\begin{equation}
\hat\chi(p,\theta,\bar\theta) = \hat\phi(p,\theta,\bar\theta) + 
g \int_{\hat\S} {\rm d}^3k\,{\rm d}^3l \, f_1(p;k,l) \delta^{(3)}(p-k-l) 
\hat\phi(k) \hat\phi(l) + O(g^2) \, ,
\label{chiordg}
\end{equation}
where $f_1(p;k,l)$ is a known function, whose exact form will not be
important for the discussion in this section. The observations at the
end of the previous subsection  would suggest that the superfield
(\ref{chiordg}) is only chiral at leading order, since the order $g$
term involves the product of chiral superfields with different
arguments. It is not difficult, however, to show that $\hat\chi$ is
actually chiral. Let us start with the truncation to order $g$. Since
$\phi$ is chiral, its Fourier transform is of the form
\begin{equation}
\hat\phi(p,\theta,\bar\theta) = {\rm e}^{\frac{1}{\sqrt{2}}p_-\theta\bar\theta}
\sum_{n=0}^4 {\hat a}^{(n)}(p)\,[\theta]^n \, .
\end{equation} 
Substituting into (\ref{chiordg}) we get
\begin{eqnarray}
&&\hsp{-0.7}\hat\chi(p,\theta,\bar\theta) = \hat\phi(p,\theta,\bar\theta) + 
g \!\sum_{m,n=0}^4\![\theta]^m[\theta]^n \!\int \!{\rm d}^3k\, {\rm d}^3l \, 
f_1(p;k,l) \delta^{(3)}(p-k-l) {\hat a}^{(m)}(k){\hat a}^{(m)}(l) 
{\rm e}^{\frac{1}{\sqrt{2}}(k_-+l_-)\theta\bar\theta} \nn \\
&&\hsp{-0.7}= \hat\phi(p,\theta,\bar\theta) + 
g \sum_{m,n=0}^4[\theta]^m[\theta]^n \int {\rm d}^2k\, {\rm d}^2l\,
\delta^{(2)}(p-k-l) \label{chiralchisteps} \\
&& \hspace*{3.8cm}\times \int \!{\rm d}k_- \, {\rm d}l_- f_1(p;k,l) 
{\hat a}^{(m)}(k) {\hat a}^{(n)}(l) \delta(p_--k_--l_-) 
{\rm e}^{\frac{1}{\sqrt{2}}(k_-+l_-)\theta\bar\theta} \nonumber \\
&&\hsp{-0.7}= \hat\phi(p,\theta,\bar\theta) + g \sum_{m,n=0}^4
[\theta]^m[\theta]^n {\rm e}^{\frac{1}{\sqrt{2}}p_-\theta\bar\theta}
\!\int \!{\rm d}^3 k\, {\rm d}^2 l \, \delta^{(2)}(p-k-l) \left.\left[
f_1(p;k,l) {\hat a}^{(m)}(k){\hat a}^{(n)}(l) \right]\right|_{l_-=p_--k_-}
\nonumber \, ,
\end{eqnarray}
where in the last step we have computed the $l_-$ integral using the 
$\delta$-function. The final expression can be rewritten as 
\begin{equation}
\hat\chi(p,\theta,\bar\theta) = {\rm e}^{\frac{1}{\sqrt{2}} 
p_-\theta\bar\theta}\sum_{n=0}^4 {\hat b}^{(n)}(p)\,[\theta]^n \, ,
\label{chichir0}
\end{equation}
with 
\begin{equation}
{\hat b}^{(n)}(p) = {\hat a}^{(n)}(p) + g \sum_{m=0}^4 
\int {\rm d}^3 k\,{\rm d}^2 l \, \delta^{(2)}(p-k-l) 
\left.\left[f_1(p;k,l) {\hat a}^{(m)}(k){\hat a}^{(n-m)}(l) 
\right]\right|_{l_-=p_--k_-} \, .
\label{chichirg}
\end{equation}
Therefore $\hat\chi$ at order $g$ is chiral, since it has the dependence
on  $\theta$ and $\bar\theta$ which characterises chiral superfields
in momentum space. However, the momentum space component fields,
${\hat b}^{(n)}(p)$, take a rather complicated form.

The generalisation to all orders is straightforward. The term of order 
$g^n$ in $\hat\chi$ is of the form
\begin{eqnarray}
&& \hspace*{-0.7cm}
g^n\!\!\sum_{m_1,\ldots,m_{n+1}=0}^4 \!\!\! [\theta]^{m_1} \cdots 
[\theta]^{m_{n+1}}
\! \int \! {\rm d}^3 k_1 \cdots {\rm d}^3 k_{n+1} \, f_n(p;k_1,\ldots,k_{n+1}) 
\,\delta^{(3)}(p-k_1-\cdots-k_{n+1}) \nonumber \\
&& \hspace*{4.1cm} \times \,
{\hat a}^{(m_1)}(k_1)\cdots{\hat a}^{(m_{n+1})}(k_{n+1}) \,
{\rm e}^{\frac{1}{\sqrt{2}} ({k_1}_-+\cdots+{k_{n+1}}_-)\theta\bar\theta} \, ,
\label{chiralchign}
\end{eqnarray}
for some function $f_n(p;k_1,\ldots,k_{n+1})$. Proceeding as in
(\ref{chiralchisteps}) we can perform the integration over
$(k_{n+1})_-$ using the $\delta$-function. This produces the correct
exponential required for the chirality of the above expression,
\begin{equation}
\left. {\rm e}^{\frac{1}{\sqrt{2}} ({k_1}_-+\cdots+{k_{n+1}}_-)
\theta\bar\theta} \right|_{{k_{n+1}}_- = p_--{k_1}_--
\cdots-{k_n}_-} ~ \longrightarrow ~
{\rm e}^{\frac{1}{\sqrt{2}} p_- \theta\bar\theta} \, .
\end{equation}
Then we can rewrite (\ref{chiralchign}) in a manifestly chiral form,
\begin{equation}
{\rm e}^{\frac{1}{\sqrt{2}} p_-\theta\bar\theta} 
\sum_{m=0}^4  {\hat b}_n^{(m)}(p) \,[\theta]^m \, ,
\end{equation}
where
\begin{eqnarray}
&& \hsp{-0.6}{\hat b}_n^{(m)}(p) = g^n \sum_{m_1,\ldots,m_n=0}^4 
\int {\rm d}^3k_1 \cdots
{\rm d}^3k_n {\rm d}^2k_{n+1} \, \delta^{(2)}(p-k_1-\cdots k_{n+1}) \\
&& \hsp{-0.6}\times
\left.\left[ f_n(p;k_1,\ldots,k_{n+1})\, {\hat a}^{(m_1)}(k_1)\cdots
{\hat a}^{(m_n)}(k_n){\hat a}^{(m-m_1-\cdots-m_n)}(k_{n+1})
\right]\right|_{{k_{n+1}}_-=p_--{k_1}_--\cdots-{k_n}_-} \, .
\nonumber
\end{eqnarray}
Therefore the redefined superfield $\hat\chi(p,\theta,\bar\theta)$ is 
indeed chiral,
\begin{equation}
{\hat d}^m_p \,\hat\chi(p,\theta,\bar\theta) = 0 \, ,
\end{equation}
although this is not obvious from the expression of $\hat\chi$ in
terms of $\hat\phi$. It is the presence of the $\delta$-function in
the field redefinition which guarantees that the $\hat\chi$ has the
correct dependence on the $\theta$ and $\bar\theta$ variables for a
chiral superfield.

We now outline the derivation of the condition (\ref{dbar chit})
satisfied by the superfield $\wtilde\chi$. In position space the
inverse field redefinition expressing $\wtilde\chi$ in terms of $\phi$
and $\bar\phi$ reads 
\be
\wtilde \chi(x,\theta) = \bar \phi(x,\theta) - 
i\frac{2}{3}g \frac{\parm}{\partial_{1-}}
\frac{\partial_{2-}\parm}{(\partial_1,\partial_2)}\,
\bar\phi(x,\theta)\,\phi(x,\theta) - i\frac{2}{3}g 
\frac{\parm}{\partial_{2-}}
\frac{\partial_{1-}\parm}{(\partial_1,\partial_2)}\,
\phi(x,\theta)\,\bar\phi(x,\theta)
+\calO(g^2),
\ee
where $(\partial_1,\partial_2) \equiv (\partial_{1-}\partial_2
- \partial_{2-}\partial_1)$.  Here, using the position space derivative 
conventions of \cite{GR}, the subscript `$1$' in $\partial_1$ denotes 
that the derivative acts on the \textit{first} superfield only, and so on
\footnote{As an example, consider the inverse transformation 
\[ 
\chi_p = \phi_p - i\frac{2}{3}\,g\int_{qr}\delta(p-q-r)\,
\frac{q_-r_-}{q_-r - r_-q}\,\phi_q\,\phi_r. 
\]
This can symbolically be expressed in position space as
\[ 
\chi(x) = \phi(x) - i\frac{2}{3}g\, \frac{\partial_{1-} \partial_{2-}}
{(\partial_1, \partial_2)}\,
\phi(x)\,\phi(x). 
\]}.
Acting with $\bar d_m$ on this
expression and using the fact that $\bar\phi$ is anti-chiral, we get
\be
\bar d_m \wtilde \chi = - i\frac{2}{3}g \frac{\parm}
{\partial_{1-}}\frac{\partial_{2-}\parm}{(\partial_1,\partial_2)}\,
\bar\phi \left(\bar d_m \phi \right) - i\frac{2}{3}g 
\frac{\parm}{\partial_{2-}} 
\frac{\partial_{1-}\parm}{(\partial_1,\partial_2)}
\left(\bar d_m \phi\right) \bar\phi +\calO(g^2) \, .
\ee
Using (\ref{con5}) we then obtain
\be
\bar d_m \wtilde \chi = - i\frac{2}{3}g \frac{i}{6\sqrt 2}
\veps_{mnpq} \frac{\parm^2}
{(\partial_1,\partial_2)} \left\{
 \frac{\partial_{2-}}{\partial_{1-}}\,
\bar\phi \left(d^n d^p d^q \frac{1}{\parm}\bar \phi \right)
 + \frac{\partial_{1-}}{\partial_{2-}} 
\left(d^n d^p d^q \frac{1}{\parm} \bar \phi\right) \bar\phi \right\}
+\calO(g^2).
\ee
Moving $\parm^2$ to the left hand side and identifying $\bar\phi$ with
$\wtilde\chi$ at zero-th order leads to the schematic form in
(\ref{dbar chit}).  We can further simplify this equation using the
definitions of the $\partial_{1-}$ and $\partial_{2-}$ operators to
obtain
\be
\bar d_m \wtilde \chi = - i\frac{2}{3}g \frac{i}{6\sqrt 2}
\veps_{mnpq} \frac{\parm^2}{(\partial_1,\partial_2)} \left\{
\left(\frac{1}{\parm}\bar\phi\right) \left(d^n d^p d^q \bar \phi \right)
 + \left(d^n d^p d^q \bar \phi\right) \left( \frac{1}{\parm} 
\bar\phi \right) \right\} + \calO(g^2).
\ee

\subsection{``Inside-out'' relations between $\chi$ and $\widetilde\chi$}
\label{insideout}

We begin with the $\widetilde\chi$ transformation in momentum space 
(see appendix \ref{app:chitilde})
\be
\label{chitildeB2}
\widetilde\chi_1 = \sum_{n=2}^{\infty}g^{n-2} \int_{2\ldots n} 
\delta(p_1-p_2-\cdots -p_n)\sum_{s=2}^{n} \frac{p_{1-}}{p_{s-}} 
\Gamma^{n,s} \chi_2 \cdots \chi_{s-1} \bar\phi_s \chi_{s+1} \cdots \chi_{n}.
\ee
Now making use of the constraint relations ${\hat d}^m\chi(p) = 0$
(see appendix \ref{constraint}) and (\ref{con2}), which in momentum 
space is ${\hat d}^4 \bar\phi(p) \sim p_-^2 \phi_p$, it is straightforward 
to see  that
\bea
\label{INOUT2}
\hat d^a\hat d^b\hat d^c\hat d^d \widetilde\chi_1 &=& 2\veps^{abcd}
\sum_{n=2}^{\infty}g^{n-2} \int_{2\ldots n}\delta(p_1-p_2-\cdots -p_n) \nn \\ 
&&\qquad\qquad  \sum_{s=2}^{n} \frac{p_{1-}}{p_{s-}}p_{s-}^2 \Gamma^
{n,s} \chi_2 \ldots \chi_{s-1} \phi_s \chi_{s+1} \ldots \chi_{n}.
\eea
We substitute for $\phi_s$ in (\ref{INOUT2}) using the expansion in 
(\ref{phi ansatz}),
\be
\label{}
\phi_1 = \sum_{n=2}^{\infty} g^{n-2} \int_{2\ldots n} 
\delta^{(3)}(p_1-p_2-\cdots-p_n) \Gamma_{1,2,\ldots,n} 
\chi_2\ldots\chi_n,
\ee
to get a relation in position space of the form 
\[
\frac{d^4}{\parm^2}\widetilde\chi \sim \chi + g\chi^2 + g^2\chi^3 
+ \cdots\,.
\]
The exact relation in position space is
\bea
\label{1e}
d^md^nd^pd^q\widetilde\chi &\!\!=\!\!& 2\,\veps^{mnpq} 
\left\{ \parm^2\chi + i\frac{2}{3}g \left[ 
\frac{\partial_{1-} \partial_{2-}}{(\partial_1,\partial_2)} 
\parm^2\left(\chi\chi\right) \right. \right. \\
&& \left. \left. \!- \frac{\parm}{\partial_{1-}}
\frac{\parm\partial_{2-}}{(\partial_1,\partial_2)}
\left(\parm^2 \chi\right)\chi- \frac{\parm}{\partial_{2-}}
\frac{\parm\partial_{1-}}{(\partial_1,\partial_2)}\chi
\left(\parm^2\chi\right) \right] + \mathcal{O}(g^2) \right\}. 
\rule{0pt}{20pt} \nn
\eea
Proceeding from (\ref{chitildeB2}) in a similar manner and making use
of (\ref {con5}), we derive
\be
\frac{d^3}{\parm}\widetilde \chi \sim \bar d \chi + g\chi \bar d 
\chi + g^2 \chi^2\bar d \chi+\cdots\,,
\ee
which is explicitly
\bea
\label{2e}
d^md^nd^p\widetilde\chi &\!\!=\!\!& -i\sqrt{2}\,\veps^{mnpq}
\left\{ \parm \bar d_q \chi  + i\frac{2}{3}g \left[ 
\frac{\partial_{1-} \partial_{2-}}
{(\partial_1,\partial_2)}
\, \parm\Big((\bar d_q\chi)\chi+\chi(\bar d_q\chi)\Big) 
\right. \right.  \\ 
&& \left. \left. \!- \frac{\parm}{\partial_{1-}}
\frac{\parm\partial_{2-}}{(\partial_1,\partial_2)}
\left(\parm\bar d_q\chi\right)\chi 
-\frac{\parm}{\partial_{2-}}
\frac{\parm\partial_{1-}}{(\partial_1,\partial_2)}\,\chi
\left(\parm\bar d_q \chi\right)\right] +  
\mathcal{O}(g^2) \right\}. \rule{0pt}{20pt} \nn
\eea
Similarly, using respectively (\ref{con4}), (\ref{con3}) and the
second relation in (\ref{con1}), we derive
\be
d^2\widetilde \chi \sim \bar d\,^2\chi + g\bar d\chi\bar d \chi 
+ g\chi\bar d\,^2\chi + g^2\chi\bar d \chi\bar d \chi 
+ g^2\chi^2\bar d\,^2\chi+\cdots\,,
\ee
\be
\parm d\widetilde \chi \sim \bar d\,^3\chi 
+ g\bar d\chi \bar d\,^2\chi + g\chi
\bar d\,^3 \chi + \cdots\,,
\ee
and 
\be
\parm^2 \widetilde \chi \sim \bar d\,^4 \chi + 
g\bar d\,^2\chi\bar d\,^2\chi + g
\bar d\chi\bar d\,^3\chi + g\chi\bar d\,^4\chi + \cdots.
\ee
It is straightforward, though tedious, to work out the coefficients at
all orders.  We will not present the details here.

\section{General form of the coefficients $\Gamma_{1,2,\ldots,n}$:
proof by induction}
\label{app:Gamma}

We want to prove that
\be
\label{gamma conjecture}
\Gamma_{+,2,\ldots,m} = \left(i \frac{2}{3}\right)^{m-2} 
\frac{ p_{2-} p_{3-}^2 p_{4-}^2 \ldots p_{(m-1)-}^2 p_{m-}}
{(p_2\,, p_3)(p_3\,, p_4)\ldots(p_{m-1}\,, p_m)}\,,
\qquad \forall m \ge 3.
\ee
This can be done by induction on $m$. The expressions in (\ref{first
gammas}) provide the initial step. We now assume that (\ref{gamma
conjecture}) is true for all $m \le n-1$. We then need to show that
(\ref{gamma conjecture}) is true for $m=n$ as well.

Substituting for the $\Gamma$'s on the r.h.s. in the recurrence
relation (\ref{recurrence}), we get
\bea
&& \hsp{-1.75}\Gamma_{1,2,\ldots,n} = -\frac{2i}{3}\fr{\omega}
\sum_{j=2}^{n-1} \left[\frac{\left\{ \left(p_2+\cdots+p_j\right),
\left( p_{j+1}+\cdots+p_n\right)
\right\}}{p_{2-}+p_{3-}+\cdots+p_{n-}} \right. \nn \\ 
 &&\left.\times \left(\frac{2i}{3}\right)^{j-2}\!\!\!\!\!
\frac{p_{2-}p_{3-}^2\cdots p_{j-}}{(p_2\,,p_3)\ldots(p_{j-1}\,,p_j)} 
\left(\frac{2i}{3}\right)^{n-j-1}\!\!\!\!\!
\frac{p_{(j+1)-}p_{(j+2)-}^2\cdots p_{n-}}{(p_{j+1}\,,p_{j+2})
\cdots (p_{n-1}\,,p_n)} \right]. \nn
\eea
Then we multiply and divide by
$\displaystyle\frac{(p_j\,,p_{j+1})}{p_{j-}p_{(j+1)-}}$ and pull the
$j$ independent factors out of the sum to obtain
\bea
&& \hsp{-1.75} \Gamma_{1,2,\ldots,n} = \left(
\frac{2i}{3}\right)^{n-2}\!\!  \frac{ p_{2-} p_{3-}^2 
\ldots p_{(n-1)-}^2 p_{n-}}{(p_2\,, p_3)(p_3\,, p_4)
\ldots(p_{n-1}\,, p_n)} \left[ \frac{-1}{\omega(p_{2-}+p_{3-}
+\cdots+p_{n-})} \right. \nn \\ 
&&\left. \times \sum_{j=2}^{n-1} \frac{
\left\{(p_2+\cdots+ p_j),(p_{j+1}+\cdots+ p_n)\right\}
(p_j\,,p_{j+1})}{p_{j-}p_{(j+1)-}}\right]. \nn
\eea
If we can show that, when $\bm{p}_1 = \bm{p}_2 + \cdots + \bm{p}_n$,
the expression within the square brackets is equal to one, the proof
is complete. It is easy to see that this is indeed the case,
\bea
&&\hsp{-3}\frac{1}{\omega(p_{2-}+p_{3-}+\cdots+p_{n-})}
\sum_{j=2}^n \frac{p_j}{p_{j-}} \left\{p_j,p_2+\cdots+p_n\right\} \nn \\
&&\hsp{-1}= \frac{1}{\omega(p_{2-}+\cdots+p_{n-})}
\left[ \frac{}{}\left(p_2 + \cdots+p_n\right)\left(
\bar p_2 +\cdots+\bar p_n\right)  \right.\nn \\
&& \hsp{0.85} \left. -\left(\frac{p_2 \bar p_2}{p_{2-}} + \cdots + 
\frac{p_n \bar p_n}{p_{n-}} \right)\left(p_{2-}+\cdots+p_{n-}
\right) \right] \nn \\
&& \hsp{-1} = \fr{\omega} \left( \omega_1 - \omega_2 - \omega_3 
-\cdots - \omega_n \right) = 1 \, ,
\eea
since $\bm{p}_1 = \bm{p}_2+\cdots+\bm{p}_n$, $\omega_i \equiv
\frac{p_i\bar p_i}{p_{i-}}$ and $\omega$ by definition is $ \left(
\omega_1 - \omega_2 - \omega_3 -\cdots - \omega_n \right)$.

\section{$\bar\phi$ to all orders}
\label{phibar}

To extract a recurrence relation, we use the fact that
\be
\label{kin term}
\Tr \int \dr^3x\, \dr\theta\; \fr{\parm}\bar \phi\; 
\parp \phi = \Tr \int \dr^3x\, \dr\theta\; 
\fr{\parm}\widetilde \chi \;\parp \chi.
\ee 
From the expansion for $\phi$ in (\ref{phi ansatz}), since all the
fields have the same $x^+$ dependence and none of the $\Gamma$
coefficients depend on $x^+$, we have 
\be
\label{delplus phi}
\parp \phi_1 = \sum_{n=2}^{\infty} g^{n-2}\sum_{r=2}^n 
\int_{2\ldots n} \delta(p_1-p_2-\cdots-p_n) \Gamma_{1,2,\ldots,n}
\chi_2\ldots \chi_{r-1}\parp\chi_r \chi_{r+1} \ldots \chi_n.
\ee
Substituting (\ref{phi bar ansatz}) and (\ref{delplus phi}) in
(\ref{kin term}), then using the cyclic property of the trace to move
$\widetilde \chi$ to the front of each string and matching the
position of $\parp\chi$ in the strings by carefully relabelling, we
arrive at the following recurrence relation
\bea
\label{cascade}
&\Xi^{N-L}_{L,L+1,\ldots,N-1,2,\ldots,L-1}& =\nn \\ \nn \\ 
\!-\!\!\!\!\!\!\!\!\!\!&\underbrace{ \sum_{k=3-L}^{N-L-1}\; 
\sum_{r=1}^{L-2} }&\!\!\!\!\! \!\!\!\!\!\!\!\!\!\!
\Xi^{r+k}_{-,N-r-k+1,N-r-k+2,\ldots,N-1,2,\ldots,L-r}\; 
\Gamma_{+,L-r+1,\ldots,N-r-k}^{}\;,\nn \\
&^{1 \le r+k \le N-L;}&  \\
&^{r\ne 1 \mbox{ \small{when} }} \nn \\
&^{k = N-L-1}& \nn
\eea
for all $N\ge 5$ and for each $L\in \{3,4,\dots,N-1\}$. Here 
$\Xi_{j,k}^1 = \delta_{j,-k}$ and $\Xi_{-,j,k}$ denotes 
$\Xi_{-j-k,j,k}$. 

In the argument of $\Xi$, in the increasing sequence
$\{(N-r-k+1),(N-r-k+2),\ldots,(N-1)\}$, if for some $(r,k)$ the first
term becomes greater than $(N-1)$, the entire sequence is to be
discarded from the argument. Then the second argument becomes
`$2$', which is next to $(N-1)$. For instance at order $g$, that is $N=5$,
and for $L=3$, $r$ is restricted to be `$1$', and when $k=0$, the
sequence becomes $\{5,6,...4\}$ (since $N-r-k+1 = 5 - 1 - 0 + 1 = 5$).
Hence the entire sequence will be discarded in the argument and
$\Xi$ will look like $\Xi^1_{-,2}$ (since $L -r = 3-1=2$). As another
example, for the same order ($N=5$) and $L=4$, when $k=0$ and $r=1$
the sequence reads $\{5,6,\ldots,4\}$ and hence is discarded. Then
$\Xi$ with arguments becomes $\Xi^1_{-,2,3}$. Also remember that
because of momentum conservation, argument `$-$' in $\Xi$ is equal to
argument `$+$' in the connected $\Gamma$. 

We iterate (\ref{cascade}) explicitly for the first few cases, to get
\bea
\Xi^1_{1,2,3} &=& -\Gamma_{-2,3,1} \, ,  \nn \\
\Xi^2_{1,2,3} &=& -\Gamma_{-3,1,2} \, ,  \nn\\
\Xi^1_{1,2,3,4} &=& -\Gamma_{-2,3,4,1} + \Gamma_{1+4,4,1}\,
\Gamma_{-2,3,1+4} \, ,  \nn\\
\Xi^2_{1,2,3,4} &=& -\Gamma_{-3,4,1,2} + \Gamma_{1+2,1,2}\,
\Gamma_{-3,4,1+2} + \Gamma_{1+4,4,1}\,\Gamma_{-3,1+4,2} \, , \nn \\
\Xi^3_{1,2,3,4} &=& -\Gamma_{-4,1,2,3} + \Gamma_{1+2,1,2}\,
\Gamma_{-4,1+2,3} \, . \nn 
\eea
These $\Xi$'s can be drastically simplified employing momentum
conservation, $\bm{p}_1 +\bm{p}_2+\cdots+\bm{p}_n = 0$, and
substituting the $\Gamma$ coefficients (\ref{gamma
conjecture}). Explicitly,
\bea
\Xi^1_{-,2,3} = \frac{p_{1-}}{p_{2-}}\,\Gamma_{+,2,3}\,;\qquad &&
\Xi^2_{-,2,3} = \frac{p_{1-}}{p_{3-}}\,\Gamma_{+,2,3}\,; \nn\\
\Xi^1_{-,2,3,4} = \frac{p_{1-}}{p_{2-}}\,\Gamma_{+,2,3,4}\,;
\qquad \Xi^2_{-,2,3,4} \!\!\!&=&\!\!\! \frac{p_{1-}}{p_{3-}}\,
\Gamma_{+,2,3,4}\,;\qquad \Xi^3_{-,2,3,4} = \frac{p_{1-}}{p_{4-}}\,
\Gamma_{+,2,3,4}. \nn
\eea
For the general term we find
\be
\label{CASCADE}
\Xi^{s-1}_{-,2,\ldots,m} = \frac{p_{1-}}{p_{s-}}\, 
\Gamma_{+,2,\ldots,m}\,,
\ee
where $m \ge 3$ and $2 \le s \le m$.  The form of the generating
functional used to define the new superfields ensures that the terms
in the old and new Lagrangians involving $\parp$ cancel each
other. This is precisely the requirement (\ref{kin term}) which we
used to express the $\Xi$ coefficients in terms of the $\Gamma$
coefficients.  In the case of pure Yang--Mills \cite{EM}, the starting
point in computing the  $\Xi$ coefficients in terms of what those
authors denote as $\Upsilon$ coefficients is also quite similar. It
then comes as no surprise that our final expression for the $\Xi$'s in
terms of the $\Gamma$'s closely matches the result in pure Yang--Mills
for the corresponding coefficients, which is~\cite{EM}
\be
\Xi^{s-1}_{1,2,\ldots,n} = -\frac{p_{s-}}{p_{1-}}\, 
\Upsilon_{1,2,\ldots,n}\,. \nn
\ee
The proof of (\ref{CASCADE}) is obtained by
induction.  We already have the initial step. We now assume that
(\ref{CASCADE}) is true for all $m \le n-1$ and all $2 \le s \le
m$. We want to show that (\ref{CASCADE}) holds for $m=n$, and all
$2\le s\le n$.

Let $N=n+2$ in (\ref{cascade}). Then
\bea
\label{cascade new}
&\Xi^{n-L+2}_{L,L+1,\ldots,n+1,2,\ldots,L-1}& =\nn \\ \nn \\ 
\!-\!\!\!\!\!\!\!\!\!\!&\underbrace{ \sum_{k=3-L}^{n-L+1}\; 
\sum_{r=1}^{L-2} }&\!\!\!\!\! \!\!\!\!\!\!\!\!\!\!
\Xi^{r+k}_{-,n-r-k+3,n-r-k+4,\ldots,n+1,2,\ldots,L-r}\; 
\Gamma_{+,L-r+1,\ldots,n-r-k+2}^{}\;,\nn \\
&^{1 \le r+k \le n-L+2;}&  \\
&^{r\ne 1 \mbox{ \small{when} }} \nn \\
&^{k = n-L+1.}& \nn
\eea
for all $n\ge 3$ and for each $L\in \{3,4,\dots,n+1\}$. We can now
substitute for $\Xi$ and $\Gamma$ on the r.h.s. to get
\bea
&& \hsp{-2}\sum \sum \frac{p_{(n-r-k+3)-}+\cdots +p_{(n+1)-}+p_{2-}
+\cdots +p_{(L-r)-}}{p_{(r+k+1)-}} \\ \nn 
&& \hsp{1} \times\,\Gamma_{+,(n-r-k+3),\ldots,(n+1),2,\ldots,L-r}  
\, \Gamma_{+,L-r+1,\ldots,n-r-k+2} \,.
\eea
Multiplying and dividing by
$\displaystyle{\frac{(p_{(L-r)},p_{(L-r+1)})}{p_{(L-r)-}p_{(L-r+1)-}}}$,
we simplify this to
\[
\sum \sum \frac{p_{(n-r-k+3)-}+\cdots +p_{(L-r)-}}{p_{(r+k+1)-}} 
\left(\xi_{(L-r+1)}-\xi_{(L-r)}\right)\, 
\Gamma_{+,(n-r-k+3),\ldots,(n+1),2,\ldots,(n-r-k+2)}\,,
\]
where $\xi_k \equiv \frac{p_k}{p_{k-}}$. We have to deal with each
value of $L$ one by one. We will prove the claim for $L=n+1$, and not
show similar proofs for other values of $L$. When we substitute
$L=n+1$ in (\ref{cascade new}), it becomes
\bea
&\Xi^1_{n+1,2,\ldots,n}& = \nn \\ \nn \\ \!-\!\!\!\!\!\!\!\!\!\!&
\underbrace{ \sum_{k=2-n}^{0}\; \sum_{r=1}^{n-1} }&\!\!\!\!\! 
\frac{p_{(n-r-k+3)-}+\cdots +p_{(L-r)-}}{p_{(r+k+1)-}} 
\left(\xi_{L-r+1}-\xi_{L-r}\right)\, \Gamma_{+,(n-r-k+3),
\ldots,(n+1),2,\ldots,(n-r-k+2)} \nn \\
&^{1 \le r+k \le 1;}&  \\
&^{r\ne 1 \mbox{ \small{when} }} \nn \\
&^{k = 0.}& \nn \\ 
&=& \hsp{-0.5} \left[ \left( \frac{p_{2-}+\cdots+p_{(n-1)-}}{p_{2-}}\right)
\left( \xi_n - \xi_{n-1}\right) + \left( \frac{p_{2-}+\cdots
+p_{(n-2)-}}{p_{2-}}\right) \left( \xi_{n-1} - \xi_{n-2}\right) 
\right.\nn \\
&& \hsp{-0.5} + \left.\left( \frac{p_{2-}}{p_{2-}}\right)\left(
\xi_3 - \xi_2\right) \right] \Gamma_{+,2,\ldots,n+1} \nn \\
& =& \hsp{-0.5} \left[ \frac{p_{2-}+\cdots+p_{(n-1)-}}{p_{2-}}\xi_n - 
\frac{p_{(n-1)-}}{p_{2-}}\xi_{n-1} - 
\frac{p_{(n-2)-}}{p_{2-}}\xi_{n-2} - \cdots - 
\frac{p_{3-}}{p_{2-}}\xi_3 - \frac{p_{2-}}{p_{2-}}\xi_2 \right] \nn \\
&=& \hsp{-0.5} \frac{1}{p_{2-}p_{n-}} \left(p_2+p_3+\cdots
+p_{n-1},p_n\right) \Gamma_{+,2,\ldots,n+1}\nn \\
&=& \hsp{-0.5} \frac{p_{(n+1)-}}{p_{2-}}\Gamma_{+,2,\ldots,n}\,,
\eea
as $\bm{p}_2+\bm{p}_3 +\cdots + \bm{p}_n+ \bm{p}_{n+1} = 0$.

\section{$\widetilde\chi$ transformation}
\label{app:chitilde}

The explicit form of the transformation (\ref{phi bar ansatz}) for
$\bar\phi$ in momentum space is given by
\bea
\label{PHIBAR}
\bar\phi_1 &=&\widetilde\chi_1 +  g\int_{23}\delta(p_1-p_2-p_3)\left
( \frac{p_{1-}}{p_{2-}}\Xi_{-1,2,3}^1\,\widetilde\chi_2\chi_3 
+ \frac{p_{1-}}{p_{3-}} \Xi_{-1,2,3}^2\, \chi_2\widetilde\chi_3 
\right) \nn \\ \nn \\
&+& g^2 \int_{234} \delta(p_1-p_2-p_3-p_4)\nn \\ 
&&\left( \frac{p_{1-}}{p_{2-}} \Xi^1_{-1,2,3,4}\, 
\widetilde\chi_2\chi_3\chi_4 + \frac{p_{1-}}{p_{3-}} 
\Xi^2_{-1,2,3,4}\, \chi_2 \widetilde\chi_3\chi_4 + 
\frac{p_{1-}}{p_{4-}} \Xi^3_
{-1,2,3,4}\, \chi_2\chi_3 \widetilde\chi_4 \right) \nn \\ \nn \\
&+& \mathcal{O}(g^3) \, .
\eea
From this we get
\bea
\label{chitilde}
\widetilde\chi_1 &=& \bar\phi_1 - 
g \int_{23} \delta(p_1-p_2-p_3) \left( \frac{p_{1-}}{p_{2-}}\Xi_
{-1,2,3}^1\,\bar\phi_2\chi_3 + \frac{p_{1-}}{p_{3-}} 
\Xi_{-1,2,3}^2\, \chi_2\bar\phi_3 \right) \nn \\ \nn \\
&-& g^2 \int_{234}\delta(p_1-p_2-p_3-p_4)\nn \\ && 
\left( \frac{p_
{1-}}{p_{2-}} \Xi^1_{-1,2,3,4}\, \bar\phi_2\chi_3\chi_4 + 
\frac{p_{1-}}{p_{3-}} \Xi^2_{-1,2,3,4}\, \chi_2 \bar\phi_3\chi_4  
+ \frac{p_{1-}}{p_{4-}} \Xi^3_
{-1,2,3,4}\, \chi_2\chi_3 \bar\phi_4 \right) \nn \\  \nn \\
&+&\!\!\! g^2\!\!\!\int_{2345}\!\!\!\!\!\! \delta(p_1-p_2-p_3) 
\delta(p_2-p_4-p_5) \nn \\ 
&&\left( \frac{p_{1-}}{p_{2-}} \frac{p_{2-}}{p_{4-}}
\Xi_{-1,2,3}^1 \Xi_{-2,4,5}^1\, \bar\phi_4\chi_5\chi_3 
+ \frac{p_{1-}}{p_{2-}}\frac{p_{2-}}{p_{5-}} 
\Xi_{-1,2,3}^1 \Xi_{-2,4,5}^2\, \chi_4\bar\phi_5\chi_3 
\right) \nn \\ \nn \\
&+&\!\!\! g^2\!\!\!\int_{2345}\!\!\!\!\!\! \delta(p_1-p_2-p_3) 
\delta(p_3-p_4-p_5) \nn \\ 
&&\left( \frac{p_{1-}}{p_{3-}} \frac{p_{3-}}{p_{4-}}
\Xi_{-1,2,3}^2 \Xi_{-3,4,5}^1\, \chi_2\bar\phi_4\chi_5 
+ \frac{p_{1-}}{p_{3-}}\frac{p_{3-}}{p_{5-}} 
\Xi_{-1,2,3}^2 \Xi_{-3,4,5}^2\, \chi_2\chi_4\bar\phi_5 \right) \nn \\ \nn \\
&+&\!\!\! \mathcal{O}(g^3) \, ,
\eea
which is easily verified order by order by substituting for $\bar\phi$
from (\ref {PHIBAR}).

\begin{figure}[!htb]
\hsp{4.7}
   \includegraphics[width=0.3\textwidth]{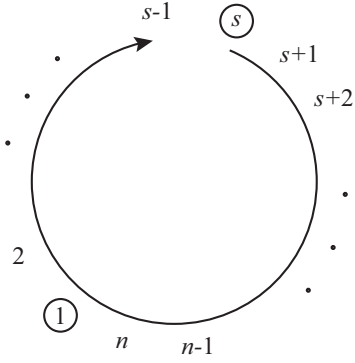}
   \caption{$\Gamma^{n,s}$ (for $n\ge 3$) - the first argument is $s$, the 
next $s+1$ and so on, increasing up to $n$ and then wrapping over to $1$, 
whence they go up to $s-1$. The circled numbers come with a minus sign in 
the argument, and $\Gamma^{2,2} = 1$. }
 \label{fig:gamma}
\end{figure}

By relabelling momenta in (\ref{chitilde}), we arrive at 
\bea
\label{chitilde2}
\widetilde\chi_1 &=& \bar\phi_1 - g\int_{23} \delta(p_1-p_2-p_3) 
\left( \frac{p_{1-}}{p_{2-}}\Xi_{-1,2,3}^1\,\bar\phi_2\chi_3 
+ \frac{p_{1-}}{p_{3-}} \Xi_{-1,2,3}^2\, \chi_2\bar\phi_3 \right) \nn \\
&-& g^2 \int_{234} \delta(p_1-p_2-p_3-p_4) \left\{
\frac{p_{1-}}{p_{2-}} \left(\Xi^1_{-1,2,3,4} - \Xi^1_{-1,2+3,4}
\Xi^1_{-(2+3),2,3}\right)\, \bar\phi_2\chi_3\chi_4 \right.\nn \\ 
&&\;\;\;\;\;\;+ \frac{p_{1-}}{p_{3-}} \left( \Xi^2_{-1,2,3,4} 
- \Xi^1_{-1,2+3,4}\Xi^2_{-(2+3),2,3}-\Xi^2_{-1,2,3+4}
\Xi^1_{-(3+4),3,4} \right)\, \chi_2 \bar\phi_3\chi_4 \nn \\ 
&&\;\;\;\;\;\;+ \left. \frac{p_{1-}}{p_{4-}} \left( \Xi^3_
{-1,2,3,4} - \Xi^2_{-1,2,3+4}\Xi^2_{-(3+4),3,4} \right)\, 
\chi_2\chi_3 \bar\phi_4 \right\} \nn \\ 
&+& \mathcal{O}(g^3).
\eea
We use the following relations from (\ref{cascade})
\bea
\Xi^1_{1,2,3} &=& -\Gamma_{-2,3,1}\;\;\;; \nn \\
\Xi^2_{1,2,3} &=& -\Gamma_{-3,1,2}\;\;\;; \nn\\
\Xi^1_{1,2,3,4} &=& -\Gamma_{-2,3,4,1} + \Gamma_{1+4,4,1}\Gamma_
{-2,3,1+4}\;\;\;; \nn\\
\Xi^2_{1,2,3,4} &=& -\Gamma_{-3,4,1,2} + \Gamma_{1+2,1,2}\Gamma_
{-3,4,1+2} + \Gamma_{1+4,4,1}\Gamma_{-3,1+4,2}\;\;\;; \nn \\
\Xi^3_{1,2,3,4} &=& -\Gamma_{-4,1,2,3} + \Gamma_{1+2,1,2}\Gamma_
{-4,1+2,3}\;\;, \nn 
\eea
to simplify (\ref{chitilde2}) to
\bea
\widetilde\chi_1 &=& \bar\phi_1 + g\int_{23} \delta(p_1-p_2-
p_3) \left( \frac{p_{1-}}{p_{2-}}\Gamma_{-2,3,-1}\,\bar\phi_2\chi_3 
+ \frac{p_{1-}}{p_{3-}} \Gamma_{-3,-1,2}\, \chi_2\bar\phi_3 \right) 
\nn \\ \nn \\
&+&\!\!\! g^2 \int_{234}\!\! \delta(p_1-p_2-p_3-p_4)\nn \\ 
&& \hsp{0.5}\times\left( \frac{p_{1-}}{p_{2-}} \Gamma_{-2,3,4,-1}\, 
\bar\phi_2\chi_3\chi_4 + \frac{p_{1-}}
{p_{3-}} \Gamma_{-3,4,-1,2}\, \chi_2 \bar\phi_3\chi_4 + 
\frac{p_{1-}}{p_{4-}} \Gamma_{-4,-1,2,3}\, \chi_2\chi_3 
\bar\phi_4 \right) \nn \\ \nn \\
&+& \!\!\!\mathcal{O}(g^3).
\eea
For the generalisation to all orders we conjecture the following form
\be
\label{conjecture}
\widetilde\chi_1 = \sum_{n=2}^{\infty}g^{n-2} \int_{2\ldots n} 
\delta(p_1-p_2-\cdots -p_n)\sum_{s=2}^{n} \frac{p_{1-}}{p_{s-}} 
\Gamma^{n,s} \chi_2 \ldots \chi_{s-1} \bar\phi_s 
\chi_{s+1} \ldots \chi_{n},
\ee
where $\Gamma^{n,s} \equiv \Gamma(.)$ and the arguments within the 
parentheses are to be filled according to the rule given in figure 
\ref{fig:gamma}. 

\section{Details of the component expansion for the superfields $\chi$
and $\wtilde\chi$ to order $g$}
\label{compdetails}

We present here some details of the calculation of the gluon vertices
in the component Lagrangian obtained from our superspace result of
section \ref{sec:MHVlag}. 

We start with the inverse transformations (\ref{invredef}) expressing
$\chi$ and $\wtilde\chi$ to order $g$ and substitute the truncated
component expansions (\ref{phi-comps}) of $\phi$ and and
$\bar\phi$. This yields the new superfields expressed in terms of the
gauge bosons as
\bea
\chi_p &=& \! \left\{ -\frac{1}{ip_-} A_p + g\!\!\int_{qr} \!\! 
\delta(p-q-r) \;\Gamma_{p,q,r} \frac{1}{q_- r_-}\; A_q\,A_r + 
\calO(g^2) \right\}  
\er^{-\frac{i}{\sqrt 2} \ttb i p_-} \nn \\ \nn \\
&-& \!\!\! \frac{1}{12} \langle \theta \rangle^4\, 
\left\{ ip_- \bar A_p + g\!\!\int_{qr} \!\!\!\!\! 
\delta(p-q-r) \; \Gamma_{p,q,r} \left( \frac{q_-}{r_-}\; 
 \bar A_q \, A_r  + \frac{r_-}{q_-}\; A_q\, \bar A_r \right) 
\!+\! \calO(g^2) \right\}, \nn
\eea
\bea
\widetilde \chi_p =&-&\!\!\! \frac{1}{ip_-} \bar A_p\; 
\er^{\frac{i}{\sqrt 2} \ttb ip_-}  
- g\int_{qr}\!\! \delta(p-q-r) \left( \frac{p_-}{q_-}\; 
\Gamma_{-q,r,-p} \frac{1}{q_- r_-} \;
\bar A_q\, A_r\; \er^{\frac{i}{\sqrt 2} \ttb i (q_- - r_-)} 
\right. \nn \\ \nn \\
&&+ \left. \frac{p_-}{r_-}\; \Gamma_{-r,-p,q} 
\frac{1}{q_- r_-}\; A_q\, \bar A_r \;  
\er^{\frac{i}{\sqrt 2} \ttb i (r_- - q_-)} \right) + \calO(g^2) \nn \\
&-&\!\!\! \frac{1}{12} \langle \bar\theta\, \rangle^4 \left\{ 
ip_- A_p - g\int_{qr}\!\! \delta(p-q-r) 
\left( \frac{p_-}{q_-}\,\Gamma_{-q,r,-p}\,
\frac{q_-}{r_-}+ \frac{p_-}{r_-}\,\Gamma_{-r,-p,q}\,
\frac{r_-}{q_-}\right) A_q\,A_r  \right. \nn \\ \nn \\
&& + \calO(g^2) \biggr\} \nn \\ \nn \\
&+&\!\!\! \frac{1}{12} \langle \theta \rangle^4 
\left\{ g\!\int_{qr}\!\! \delta(p-q-r)
\left( \frac{p_-}{q_-}\,\Gamma_{-q,r,-p}\, \frac{r_-}{q_-}+ 
\frac{p_-}{r_-}\,\Gamma_{-r,-p,q}\,\frac{q_-}{r_-} 
\right) \bar A_q\, \bar A_r + \calO(g^2)
 \right\} \nn \\ \nn \\
 &-&\!\!\! \frac{1}{144} \langle \theta \rangle^4 
\langle \bar \theta \, \rangle^4
 \left\{ g\!\int_{qr}\!\! \delta(p-q-r) \left( 
\frac{p_-}{r_-}\,\Gamma_{-r.-p,q}\, 
 q_-r_-\; \bar A_q\, A_r \right.\right.\nn \\ \nn \\ 
&&+ \left. \frac{p_-}{q_-}\,\Gamma_{-q,r,-p}\, 
 q_- r_- \; A_q\, \bar A_r \right) + \calO(g^2) 
 \biggr\}\;. \nn
\eea
Substituting the $\Gamma$ coefficients  derived in section
\ref{sec:phi}, $\Gamma_{+,q,r} = \frac{q_- r_- }{(q,r)}$, we get
\bea
\label{chiinA}
\hsp{-0.5}\chi_p &=& \! \left\{ -\frac{1}{ip_-} A_p + i\frac{2}{3}g\!\!
\int_{qr} \!\! \delta(p-q-r) \;
\frac{1}{(q,r)}\; A_q\,A_r + \calO(g^2) \right\}  
\er^{-\frac{i}{\sqrt 2} \ttb i p_-} \nn \\ \nn \\
&-& \!\!\! \frac{1}{12} \langle \theta \rangle^4 
\left\{ ip_- \bar A_p + i\frac{2}{3}g\!\!\int_{qr} \!\!\!\!\! 
\delta(p-q-r)\!\left( \frac{q_-^2}{(q,r)}\; 
 \bar A_q \, A_r  + \frac{r_-^2}{(q,r)}\; A_q\, \bar A_r \right) 
\!\!+ \calO(g^2) \right\},
\eea
\bea
\label{chitinA}
\hsp{-0.5} \widetilde \chi_p = &-&\!\!\! \frac{1}{ip_-} \bar A_p\; 
\er^{\frac{i}{\sqrt 2} \ttb ip_-}  
+ i\frac{2}{3}g\int_{qr} \delta(p-q-r)\; \left( 
\frac{p_-^2}{q_-^2}\; \frac{1}{(q,r)} \;
\bar A_q\, A_r\; \er^{\frac{i}{\sqrt 2} \ttb i (q_- - r_-)} 
\right. \nn \\ \nn \\
&& +\left. \frac{p_-^2}{r_-^2}\; \frac{1}{(q,r)}\; A_q\, \bar A_r \;  
\er^{\frac{i}{\sqrt 2} \ttb i (r_- - q_-)} \right) 
+ \calO(g^2) \nn \\ \nn \\
&-& \!\!\! \frac{1}{12} \langle \bar\theta\, \rangle^4 
\left\{ ip_- A_p  + i\frac{4}{3}g\int_{qr}\!\! \delta(p-q-r)\; 
\frac{p_-^2}{(q,r)}\; A_q\,A_r 
+ \calO(g^2) \right\} \nn \\ \nn \\
&-&\!\!\! \frac{1}{12} \langle \theta \rangle^4 \left\{ 
i\frac{2}{3}g\!\int_{qr}\!\! \delta(p-q-r)\; 
\frac{p_-^2}{(q,r)}\left( \frac{q_-^2}{r_-^2} + 
\frac{r_-^2}{q_-^2}\right) \bar A_q\, \bar A_r + \calO(g^2)\right\} 
\nn \\ \nn \\
 &+&\!\!\! \frac{1}{144} \langle \theta \rangle^4 
\langle \bar \theta \, \rangle^4
 \!\left\{ i\frac{2}{3}g\!\int_{qr}\!\!\! \delta(p-q-r)\! 
\left( \frac{p_-^2 q_-^2}{(q,r)}\; 
 \bar A_q\, A_r + \frac{p_-^2 r_-^2}{(q,r)} \; A_q\, 
\bar A_r \!\right) \!+\! \calO(g^2)
 \right\}\,.
\eea
To show that the new Lagrangian in terms of $\chi$ and
$\widetilde\chi$  reduces to the MHV Lagrangian for pure Yang--Mills
derived in~\cite{GR,PM}, we need to rewrite the   new superfields in
terms of the redefined component fields $B$ and $\widetilde B$,  which
make up the MHV Lagrangian for Yang--Mills. For this purpose we make
use of the explicit field redefinitions derived  in~\cite{EM},
which we reproduce here
\bea
A_p &=& B_p - g\int_{qr} \delta(p-q-r)\; \frac{p_-}{(q,r)}\; 
B_q\,B_r + \calO(g^2)\,,\\
\bar A_p &=& \widetilde B_p - g\int_{qr} \delta(p-q-r)\;  
\left( \frac{q_-^2}{p_-} 
\frac{1}{(q,r)}\; \widetilde B_q\,B_r + \frac{r_-^2}{p_-}
\frac{1}{(q,r)}\; B_q \widetilde B_r \right)
+ \calO(g^2)\,.
\eea
Substituting these in (\ref{chiinA}) and (\ref{chitinA}), we obtain
\bea
\label{chiinB}
\hsp{-0.5} \chi_p &=& \! \left\{ -\frac{1}{ip_-} B_p - \frac{i}{3}g\!\!
\int_{qr} \!\! \delta(p-q-r) \;
\frac{1}{(q,r)}\; B_q\,B_r + \calO(g^2) \right\}  
\er^{-\frac{i}{\sqrt 2} \ttb i p_-} \nn \\ \nn \\
&-& \!\!\! \frac{1}{12} \langle \theta \rangle^4\, 
\left\{ ip_- \widetilde B_p
 - \frac{i}{3}g\!\!\int_{qr} \!\!\!\!\! \delta(p-q-r)
\!\left( \frac{q_-^2}{(q,r)}\; 
 \widetilde B_q \, B_r  + \frac{r_-^2}{(q,r)}\; B_q\, 
\widetilde B_r \right) \!\!+ \calO(g^2) \right\},
\eea
\bea
\label{chitinB}
\hsp{-0.5}
\widetilde \chi_p &=& \left\{ -\frac{1}{ip_-} \widetilde B_p  
-ig\int_{qr}\!\!\delta(p-q-r)\;\left( \frac{q_-^2}{p_-^2}
\frac{1}{(q,r)}\; \widetilde B_q\, B_r
+ \frac{r_-^2}{p_-^2}\frac{1}{(q,r)}\; B_q \,\widetilde B_r
\right) \right.\nn \\ \nn \\ 
&& +\calO(g^2) \biggr\} \er^{\frac{i}{\sqrt 2} \ttb ip_-} \nn \\ \nn \\
&+&\!\!\! i\frac{2}{3}g\int_{qr} \delta(p-q-r)\; \left( 
\frac{p_-^2}{q_-^2}\; \frac{1}{(q,r)} \;
\widetilde B_q\, B_r\; \er^{\frac{i}{\sqrt 2} \ttb i (q_- - r_-)} 
\right. \nn \\ \nn \\
&& + \left. \frac{p_-^2}{r_-^2}\; \frac{1}{(q,r)}\; B_q\, 
\widetilde B_r \;  
\er^{\frac{i}{\sqrt 2} \ttb i (r_- - q_-)} \right) + 
\calO(g^2) \nn \\ \nn \\
&-&\!\!\! \frac{1}{12} \langle \bar\theta\, \rangle^4 
\left\{ ip_- B_p  + \frac{i}{3}g\int_{qr}\!\! \delta(p-q-r)\; 
\frac{p_-^2}{(q,r)}\; B_q\,B_r 
+ \calO(g^2) \right\} \nn \\ \nn \\
&-&\!\!\! \frac{1}{12} \langle \theta \rangle^4 \left\{ 
i\frac{2}{3}g\!\int_{qr}\!\! \delta(p-q-r)\; 
\frac{p_-^2}{(q,r)}\left( \frac{q_-^2}{r_-^2} + 
\frac{r_-^2}{q_-^2}\right) 
\widetilde B_q\, \widetilde B_r + \calO(g^2)\right\} \nn \\ \nn \\
 &+&\!\!\! \frac{1}{144} \langle \theta \rangle^4 \langle 
\bar \theta \, \rangle^4
 \left\{ i\frac{2}{3}g\!\int_{qr}\!\! \delta(p-q-r)\; 
\left( \frac{p_-^2 q_-^2}{(q,r)}\; 
 \widetilde B_q\, B_r + \frac{p_-^2 r_-^2}{(q,r)} \; B_q\, 
\widetilde B_r \right) + \calO(g^2) \right\}.
\eea
Substituting (\ref{chiinB}) and (\ref{chitinB}) in the MHV Lagrangian
presented in section \ref{sec:MHVlag} and computing the
Grassmann integrals,  we reproduce the MHV Lagrangian for pure
Yang--Mills.

\end{document}